\documentclass[12pt]{article}
\usepackage[centertags]{amsmath}
\usepackage{hyperref}
\usepackage{color}
\usepackage{amsfonts}
\usepackage{amssymb}
\usepackage{amsthm}
\usepackage{cite}
\usepackage{fullpage}
\usepackage{graphicx}
\usepackage{pgf,pgfarrows,pgfnodes}
\usepackage{graphics}
\usepackage{tikz}

\newcommand{\res}{\mathrm{Res}}

\let\chir=\bar

\let\a=u          
         
\let\i=\iota

         \let\w=\omega

\newcommand{\R}{{\mathbb R}}

\newcommand{\PP}{{\mathbb P}}
\newcommand{\be}{\begin{equation}}
\newcommand{\eeq}{\end{equation}}
\newcommand{\bea}{\begin{eqnarray}}
\newcommand{\eea}{\end{eqnarray}}
\newcommand{\ba}{\begin{array}}
\newcommand{\ea}{\end{array}}
\def\nn{\nonumber}
\newcommand{\ft}[2]{{\textstyle\frac{#1}{#2}}}

\newcommand{\ee}{\end{equation} }

\newcommand{\Tr}{\mathrm{Tr}\,}

\newcommand{\cN}{{\mathcal{N}}}

\newcommand{\cQ}{{\mathcal{Q}}}

\newcommand{\one}{{\rm 1\kern -.9mm l}}

\newdimen\tableauside\tableauside=1.0ex
\newdimen\tableaurule\tableaurule=0.4pt
\newdimen\tableaustep

\def\phantomhrule#1{\hbox{\vbox to0pt{\hrule height\tableaurule
width#1\vss}}}
\def\phantomvrule#1{\vbox{\hbox to0pt{\vrule width\tableaurule
height#1\hss}}}
\def\sqr{\vbox{%
 \phantomhrule\tableaustep
\hbox{\phantomvrule\tableaustep\kern\tableaustep\phantomvrule\tableaustep}%
 \hbox{\vbox{\phantomhrule\tableauside}\kern-\tableaurule}}}
\def\squares#1{\hbox{\count0=#1\noindent\loop\sqr
 \advance\count0 by-1 \ifnum\count0>0\repeat}}
\def\tableau#1{\vcenter{\offinterlineskip
 \tableaustep=\tableauside\advance\tableaustep by-\tableaurule
 \kern\normallineskip\hbox
   {\kern\normallineskip\vbox
     {\gettableau#1 0 }%
    \kern\normallineskip\kern\tableaurule}%
 \kern\normallineskip\kern\tableaurule}}
\def\gettableau#1 {\ifnum#1=0\let\next=\null\else
 \squares{#1}\let\next=\gettableau\fi\next}
\newtheorem{remark}{Remark}

\newcommand*\diff{\mathop{}\!\mathrm{d}}
\newcommand{\sign}{\mathrm{sign}}

\begin{document}

\font\cmss=cmss10 \font\cmsss=cmss10 at 7pt

\vskip -0.5cm

\vskip .7 cm

\hfill
\vspace{18pt}
\begin{center}
{\Large \textbf{   Gauge theories on compact toric manifolds  }}
\end{center}

\vspace{6pt}
\begin{center}	
	{\textsl Giulio Bonelli$^{\ddagger}$\footnote{\scriptsize \tt bonelli@sissa.it},
	Francesco Fucito$^{\dagger}$ \footnote{\scriptsize \tt francesco.fucito@roma2.infn.it},
	Jose Francisco Morales $^{\dagger }$ \footnote{\scriptsize \tt francisco.morales@roma2.infn.it},
	Massimiliano Ronzani \footnote{\scriptsize \tt ronzani.massimiliano@gmail.com},
	Ekaterina Sysoeva$^{\ddagger}$ \footnote{\scriptsize \tt esysoeva@sissa.it},
	Alessandro Tanzini$^{\ddagger}$ \footnote{\scriptsize \tt tanzini@sissa.it}}\\
\vspace{1cm}				
$\dagger$\textit{\small I.N.F.N. Sezione di Roma ``TorVergata''\\  Dipartimento di Fisica, Universit\`a di Roma ``TorVergata'', \\ Via della Ricerca Scientifica, 00133 Roma, Italy }\\
\vspace{6pt}
$\ddagger$\textit{\small SISSA, Via Bonomea 265, 34136 Trieste, Italy\\  I.N.F.N, Sezione di Trieste,\\I.G.A.P., Via Beirut 4, 34100 Trieste, Italy}
\vspace{6pt}\end{center}

\begin{center}
\textbf{Abstract}
\end{center}

\vspace{4pt} {\small

\noindent We compute the ${\cal N}=2$ supersymmetric partition function of a gauge theory on a four-dimensional compact toric manifold via equivariant localization.
The result is given by a piecewise constant function of the K\"ahler form with jumps along the walls where the gauge symmetry gets enhanced. The
partition function on such manifolds is written as a sum over the residues of a product of partition functions on $\mathbb{C}^2$.
The evaluation of these residues is greatly simplified by
using an ``abstruse duality'' that relates the residues at the poles of the one-loop and instanton parts of the $\mathbb{C}^2$ partition function.
As particular cases, our formulae compute the $SU(2)$ and $SU(3)$ {\it equivariant} Donaldson invariants of $\mathbb{P}^2$ and $\mathbb{F}_n$
and in the non-equivariant limit reproduce the results obtained via wall-crossing
and blow up methods in the $SU(2)$ case.
Finally, we show that the $U(1)$ self-dual connections induce an anomalous dependence on the gauge coupling,
which turns out to satisfy a $\mathcal{N}=2$ analog of the $\mathcal{N}=4$ holomorphic anomaly equations.
}
\newpage

\tableofcontents

\section{Introduction, summary and open questions}
\label{section1}

The study of $\cN=2$ supersymmetric Yang-Mills gauge theories in four dimensions (SYM) led to many interesting and deep results which opened a new perspective in the understanding of non-perturbative effects in gauge theories \cite{Seiberg:1994rs, Seiberg:1994aj}.

Major progresses in this respect have been obtained by making use of
{\it equivariant localization} for the gauge theory in the so-called $\Omega$-background \cite{Nekrasov:2002qd}.
This allowed for the exact computation of the supersymmetric path integral and low energy effective action on $\R^4$
\cite{Nekrasov:2002qd, Bruzzo:2002xf, Flume:2002az} and its orbifolds
by a discrete group
$\R^4/\Gamma$ \cite{Fucito:2004ry,Bonelli:2011jx,Bonelli:2011kv,Bonelli:2012ny,Bruzzo:2013nba,Bruzzo:2013daa}.
The first explicit computation  of the gauge theory partition function on a compact manifold, using equivariant localization, was performed in \cite{Pestun:2007rz}  on the four sphere $S^4$ and extended to the squashed case in \cite{Hama:2012bg}.
These results were generalised to some non-orientable manifolds in \cite{Bawane:2017gjf,LeFloch:2017lbt}.
The extension of these results to toric compact manifolds
was prompted by \cite{Nekrasov:2003viNO}, and the $S^2\times S^2$ and $\PP^2$ cases were considered in \cite{Bawane:2014uka,Bershtein:2015xfa,Bershtein:2016mxz}. The gauge partition function was formally written as a contour integral picking a specific set of poles  associated to (semi-)stable gauge bundles. For recent results see also \cite{Rodriguez-Gomez:2014eza,Beaujard:2020sgs}.

In this paper, building on the above mentioned results, we propose a formula valid for arbitrary {\it compact} toric manifolds and arbitrary rank of the gauge group
for the topologically twisted theory.  The computation of the gauge theory partition function on compact manifolds presents a main additional difficulty with respect to the non compact case, namely one has to perform an integration over the Coulomb branch parameters, which are in this case integrable zero modes of the dynamical fields.
After the topological twist, the ${\cal N}=2$ gauge theory turns into a
 cohomological field theory \cite{Witten:1988ze} and therefore the correlators of BPS protected observables are expected to be independent both on the metric and on the gauge coupling. Indeed this is not trivial in cohomology, because of boundary effects in the field space.
As we will show the integration over the zero modes induces an anomalous dependence in these parameters having two related important
consequences: the first is to produce some constraints on the sum over the fluxes of the gauge field, which in turn induces a non-trivial wall-crossing behaviour of the partition function. The second is that the partition function acquires
an anomalous dependence on the gauge coupling which can be characterised in terms of a holomorphic anomaly equation.
Indeed, the Coulomb moduli space over which we integrate is non compact and
this induces an anomaly arising from the boundary term.
Moreover,  in the topologically twisted ${\cal N}=2$ theory the coupling appears in the gauge fixing term which is metric dependent.
This in turn implies an induced anomalous dependence of the partition function on the metric of the manifold and the related wall crossing behaviour.

A careful analysis of the zero modes sector gives a prescription for the computation of the partition function.
After gauge fixing the Weyl symmetry, the partition function is written as a sum over the residues characterised by an ordering of the fluxes of the gauge field along
the  K\"ahler two-form. This ordering is crucial in selecting the relevant contributions  \footnote{Notice that on non compact manifolds, like ALE spaces, the sum over all fluxes is unconstrained since the Weyl symmetry is explicitly broken by the choice of the scalar vev at infinity. }.  The integral over the zero modes turns out to be ill defined at the walls of marginal stability where two (or more) fluxes coincide.
The resulting partition function is therefore piece-wise dependent on the choice of  K\"ahler two-form, with jumps at the walls where two (or more) fluxes of the gauge field curvature along the K\"ahler two-form get equal.

On the mathematical side,  gauge theory correlators compute the Donaldson invariants of the four manifold \cite{Witten:1988ze}. More precisely, the supersymmetric partition function in the $\Omega$-background computes {\it equivariant } Donaldson invariants \cite{Gottsche:2006tn}. The above mentioned jumps of the gauge theory partition function correspond to
 the well known wall-crossing behaviour in Donaldson theory, which in the framework of algebraic geometric is induced by changes in the stability conditions of the sheaves. The sum over gauge theory fluxes is shown to properly select the (semi-)stable equivariant sheaves and to nicely reproduce their topological classification \cite{Klyachko2, Kool}. The results we obtain for the partition function are tested against existing results in the mathematical literature for the $SU(2)$ case,
 based on wall crossing and blow up formulae. Some new explicit predictions for the $SU(3)$ case on $\mathbb{P}^2$ will also be given.

The anomalous dependence on the gauge coupling discussed above
is expected to be the UV ancestor of the holomorphic anomaly equations in the IR, closely connected to analogous results first found in string theory models \cite{Bershadsky:1993ta} then in the case of the computation of the partition function of the
$\mathcal{N}=4$ SYM \cite{Vafa:1994tf} and in the Donaldson invariants generating functions \cite{Malmendier:2012zz,Griffin:2012kw,Korpas:2017qdo, Korpas:2019cwg}. More recently, the derivation of the holomorphic anomaly for the twisted version of the $\mathcal{N}=4$ SYM and its relation with the mock modular forms has been discussed in \cite{Dabholkar:2020fde}. It would be also interesting to extend  our approach to the topologically twisted theories considered in \cite{Festuccia:2018rew,Festuccia:2020xtv} which generically localize both to point-like instantons and anti-instantons configurations.

In this paper we mainly focus on the
holomorphically decoupled sector of the theory and rely on
 equivariant localization. The path integral computing the partition function of a gauge theory on
 a toric manifold  localizes on the fixed points of the torus action, namely on point like instantons sitting at the origin of each toric patch covering the manifold. The path integral is computed in terms of the residues of a product of partition functions, one for each toric patch. The residues are taken at the fixed points of the torus action and are specified by the fluxes of the gauge field along the Cartan subalgebra.
 To compute such residues we use a surprising ``duality'' relation between the residues
 computed at the poles of the one-loop and instanton part of the partition function.
 This duality is rooted in the so called AGT correspondence \cite{Alday:2009aq} which connects the partition functions
 of the $\cN=2$ class $S$ theories to the conformal blocks of a two dimensional conformal field theory (CFT).
In the case of $SU(2)$ SYM such ``duality''  between the residues is a direct consequence of the Zamolodchikov recursion relations for the conformal
blocks \cite{Poghossian:2009mk}. In this paper we will present a generalization of such relation, valid for higher rank unitary gauge groups.
Once again the gauge theory quantities can be put in correspondence with the two-dimensional CFT ones: the poles of the one-loop and instanton
partition functions can, in fact, be put in correspondence with the conformal dimensions of the null states of
the reducible Verma module and the roots of the associated Kac determinant\cite{Watts:1989bn}\footnote{This generalization has been obtained in collaboration with R.Poghossian in an unpublished work. Later it has been applied to the SYM with gauge groups of rank two in \cite{Poghossian:2017atl}.}.

{\vskip .5cm}

 There are several open questions and aspects to be further analyzed, let us mention some of them. It would be interesting
to extend the present computations to gauge theories with fundamental and adjoint matter fields and perform a more thorough analysis of the higher rank cases. The latter point would provide new results for the Donaldson invariants
in the higher rank case for which very few results are known at the moment, with the notable exception of \cite{Marino:1998bm,2017arXiv170100571D}. Moreover, it would be useful for a large $N$ analysis and for the study of holography for compact manifolds.
Also, the insertion of defect operators would be an interesting aspect to investigate.

The results obtained in this paper are based on equivariant localization on the microscopic UV Lagrangian in the $\Omega$-background. It would be very interesting to study the limit of vanishing $\Omega$-background in order to establish a
connection with the integration over the $u$-plane \cite{Moore:1997pc} which is based on an IR analysis using the abelian effective gauge theory. This would possibly allow to make manifest the duality properties of the partition function and in particular to connect our results on the $\mathcal {N}=2$ holomorphic anomaly
to the theory of mock modular forms.

Further directions concern the uplift of our results to five and six dimensional gauge theories. In the case of the product manifolds $M\times S^1$ and $M\times T^2$ this  would correspond on the mathematical side to K-theoretic and elliptic Donaldson invariants respectively. More generally, one could try to extend the gluing techniques exploited in this paper to toric manifolds in higher dimensions.

{\vskip .5cm}

The paper is organized as follows:
in section \ref{section2}, we  use the localization method to derive a formula for the gauge partition function on a compact manifolds.
We describe the integration over the zero modes and the holomorphic anomaly in the gauge coupling.
In section \ref{section3} we discuss the properties of the localized partition function which will be relevant for the computations of
the following sections and in particular we discuss a remarkable duality between the perturbative and the instanton part of the partition function which will greatly simplify our computations.
In section \ref{section4} we specialise the findings of the previous sections to the case of toric manifolds.
In section \ref{section5} and  \ref{section6} we analyse the cases of $SU(2)$ gauge theory on $\PP^2$ and $\mathbb{F}_n$ 
manifolds. In section \ref{sectionsun} we present some preliminary results for the $SU(3)$ partition function on $\PP^2$.
Finally, several useful results are collected in the Appendices.

\section{Localization on compact manifolds}
\label{section2}
\setcounter{equation}{0}
 It is well known that the $\cN=2$ SYM with gauge group $U(N)$ can be formulated on any differentiable Riemaniann four-manifold by making use of twisted supersymmetry \cite{Witten:1988ze}. The bosonic content of the $\cN=2$ gauge supermultiplet includes a gauge vector $A$, a complex scalar $\Phi$ and a self-dual two-form $B^+$ which is an auxiliary field. The fermionic components
 are a one-form $\Psi$, a scalar $\eta$ and a self-dual two-form $\chi^+$. Fields are paired by a scalar supersymmetry charge $Q$.

   \subsection{Equivariant localization}

     The supercharge $Q$ can be viewed as an equivariant derivative acting on the supermanifold with coordinates the fields $\mathfrak M=(A,\bar \Phi,\chi^+)$ and
equivariant differentials $\mathfrak d\mathfrak M=(\Psi,\eta,B^+)$.  The supersymmetry action can be further deformed using an isometry $\delta_V$ of $M$
\be
 {\cal Q}= Q+ i \,\iota_V =\mathfrak d +  i \,\iota_{\Phi} + i \,\iota_{V}
 \qquad ,\qquad  {\cal Q}^2=\delta_\Phi+\delta_V
 \ee
 where $\iota_V$,  $\iota_\Phi$  is the contraction along the vector field $V$ and $\Phi$, i.e.
 $ i \,\iota_{\Phi} \mathfrak d\mathfrak M=\delta_{\Phi} \mathfrak M$,
 $ i \,\iota_{V} \mathfrak d\mathfrak M=\delta_{V} \mathfrak M$.
  The $V$-deformation represents the $\Omega$-deformation of the theory.
  In the toric case, it is
  a $U(1)^2$ local Lorentz rotation which, for each toric patch, is the local
  $\Omega$-background on ${\mathbb C}^2$ with the appropriate choice of weights
  (See Section \ref{section4} for details).
%
 The scalar supercharge action is
\be\label{eq:newSUSY1}
\begin{aligned}
&\cQ A=\Psi, &\quad &\cQ\Psi= i \,\iota_V F + D\Phi, &\quad& \cQ \Phi= i\,\iota_V\Psi,\\
&\cQ\chir\Phi=\eta, &\quad &\cQ\eta= i \iota_VD\chir\Phi+ [\Phi,\chir\Phi], \\
&\cQ\chi^+=B^+, &\quad &\cQ B^+=i\, \mathcal{L}_V\chi^++ [\Phi,\chi^+],
\end{aligned}
\ee
with $ \mathcal{L}_V=  (D\iota_V+\iota_V D) = -{\rm i} \delta_V$ the covariant Lie derivative
 associated to the action of the vector field V and
  $D=d+ [A,\cdot]$ the covariant derivative.
Notice that the consistency of the last line implies that the self-duality of the two forms is preserved by the $V$-action, namely ${\cal L}_V \star = \star {\cal L}_V$. This is equivalent to the statement that $V$ generates  an isometry of the four manifold.

The ${\cal N}=2$ SYM Lagrangian can be written
\be\label{lagrangian}
{\cal L}=\frac{i\tau}{8\pi}\Tr \left(F\wedge F\right)
+\cQ \mathcal{V}
\ee
where $\tau$
is the complex coupling, the trace is in the fundamental representation and
 \bea\label{GFFform}
\mathcal{V}=\text{Tr}\big[
i\chi^{+}\wedge F^{+}
+\frac{g^2}{4} \chi^{+}\wedge B^{+}  +
         \Psi  \wedge\star(\cQ\Psi)^\dagger +\eta\wedge\star(\cQ\eta)^\dagger\,
          \big].
\eea
Consistently, the action is gauge and ${\cal L}_V$ invariant.
In \eqref{GFFform}, after integrating out the auxiliary field $B^+$, one
gets
\be\label{ttb}
{\cal L}=\frac{i\tau}{8\pi}\Tr \left(F^-\wedge F^-\right)
+
\frac{i\bar\tau}{8\pi}\Tr \left(F^+\wedge F^+\right)
+\ldots
\ee
where $\tau=\frac{\theta}{2\pi}+i\frac{4\pi}{g^2}$.
In the topological theory $\tau$ and $\bar\tau$ are independent parameters.
Indeed, at
the quantum level, the real part of $\tau$ is not a physical parameter, since it can be absorbed by an anomalous $U(1)_R$ transformation.



\paragraph{SUSY fixed points:}
\noindent
  The path integral of the deformed gauge theory localizes around the fixed locus
  of the supersymmetry.
To identify the set of fixed points of the twisted supersymmetry \eqref{eq:newSUSY1}, we start by setting as usual all fermions to zero and then we impose the vanishing of their supersymmetric transformations
\be
\begin{aligned}\label{susyB}
  i \,\iota_V F + D\Phi = 0 \, ,\quad
  i \iota_VD\chir\Phi+ [\Phi,\chir\Phi] = 0\, .
\end{aligned}
\ee
By applying $\iota_V$ to the first equation and using $\iota_V^2=0$
and the reality condition $\Phi^\dagger=\bar\Phi$
one finds
\be
   \i_V D\bar \Phi =0 \,  \label{qpsi}.
\ee
Using (\ref{qpsi}) into (\ref{susyB}), we get
\be
[\Phi,\bar\Phi] =0
\label{pote}
\ee
which projects the field $\Phi$ onto the Cartan subalgebra.
Finally acting with the covariant derivative on the l.h.s. of (\ref{susyB}) one finds
\be
{\cal L}_V F =[F,\Phi]=0
\ee
where in the last equation we used the reality of $F$ and (\ref{pote}).
At the supersymmetric fixed points one then finds for the equivariant gauge field
\be
\mathfrak{F}\big|_{\rm f.p.}=\left( F+\Phi \right)_{\rm f.p.} ={\rm diag} \{
\mathfrak{F}_1, \ldots \mathfrak{F}_{N} \},
\label{equi}
\ee
where
\be\label{fp}
\mathfrak{F}_u= \mathfrak{F}_{u}^{\rm point} + \mathfrak{F}_{u}^{\rm flux}
\, , \,\, {\rm where}\,\, \mathfrak{F}_{u}^{\rm flux}=
2\pi \sum_{\ell=1}^\chi k^\ell_u w_\ell \, .
\ee
In \eqref{fp} $\{ w_\ell \}$, $\ell=1,\ldots \chi$ is a basis of $H^2_V(M;\mathbb{Z})$, $\chi$ being the Euler characteristic of the manifold $M$.
The first term in \eqref{fp} describes point-like instantons sitting at the fixed points of the toric action, while the second describes the equivariant fluxes of the gauge field.

\paragraph{BPS observables:}
The BPS observables of the topologically twisted
gauge theory are built by the equivariant version of the usual descent equations \cite{Bershtein:2015xfa}. The supersymmetry transformations
\eqref{eq:newSUSY1} can be succinctly rewritten
as the equivariant Bianchi identity \cite{Baulieu:2005bs}
\be
\mathfrak{ D F}\equiv
\left(-\cQ + D +i\iota_V\right)\left(F+\Psi+\Phi\right)=0\,.
\ee
for the equivariant curvature
\be
\mathfrak{F}=F+\Psi+\Phi
\ee
Indeed, \eqref{eq:newSUSY1}
can be obtained from the above
by expanding in the de Rham form degree.

In these variables the supersymmetric action takes the simple form
\be
S_{top} ={{\rm i} \tau \over 8\pi } \int_M {\rm tr}\, \mathfrak{F}^2
\ee
 More generally, one can consider the intersections of the equivariant forms built out of $\mathfrak{F}$
with elements of the equivariant cohomology
of the manifold, ${\cal T}\in H^\bullet_V(M)$ as
\be
 S_{top, sources} \equiv
\int_M {\cal T} \wedge {\cal P}(\mathfrak{F} ).
\ee
 with ${\cal P}$ a gauge invariant function of $\mathfrak{F}$.
 In the following we will discuss the case in which ${\cal P}$ is a quadratic polynomial, while for compact toric manifolds the source ${\cal T}$
reads
  \be
 {\cal T}=2\pi {\rm i} \,\tau +x^\ell w_\ell +x^{\ell \ell' } w_\ell\wedge w_{\ell'} +T\, .
 \label{donaldsonsource}\ee
 Moreover, $T$ is an arbitrary  polynomial of order two in the equivariant parameters and
 $X=\{ x^\ell$, $x^{\ell \ell' } \}$  are the variables of the Donaldson polynomials. Actually,
these generate the equivariant extension of surface and local observables which are the relevant ones for the computation of Donaldson polynomials.
Let us remark that the set of equivariant observables is richer than the non-equivariant ones and the equivariant Donaldson polynomials give a finer characterization of the differentiable manifolds.
From the quantum field theory view point, this is because the $\Omega$-background probes the gauge theory revealing a finer
BPS structure.

\subsection{Integrating around fixed points}

The computation of the partition function proceeds by integrating the zero modes around
the supersymmetric fixed points \eqref{fp} and then by using the localization formulae.


At the origins of the toric patches, which are fixed points under the toric action,
 the supersymmetry equations become exactly
those of point like instantons sitting at the origin of $\mathbb{C}^2$ after the replacement
$a\to a^\ell$, $\epsilon_a \to \epsilon_a^\ell$ and $q \to q^\ell$.
Therefore the contribution of
$\mathfrak{F}_{\rm point}$ to the gauge theory partition function factorises in a product of local factors, each of them being
given  by the partition function $Z_{\mathbb{C}^2} (a_\alpha^\ell ,\epsilon_1^\ell,\epsilon_2^\ell,q^\ell)$ with
scalar vevs $a_\alpha^\ell$, equivariant parameters $\epsilon^\ell$ and  gauge couplings
$q^\ell$ determined by the non-trivial gluing of the charts. 
The expansion of the scala vevs is along the Cartan generators
$h_\alpha$'s, $\alpha=1,\ldots,N-1$ are the generators of the Cartan subgroup of the gauge group, satisfying
\be\Tr (h^\alpha h^\beta)=G^{\alpha\beta}
\label{cartannormaliztion}\ee
where $G_{\alpha\beta}$ is the Lie algebra Cartan matrix.

The couplings $q^\ell=q\, e^{\Omega_\ell}$, where ${\Omega_\ell}$ is the the zero form part of ${\cal T}$ evaluated at the fixed point
in chart $\ell$, take properly into account the contribution of surface and local observables.

The equivariant parameters $\epsilon_a^\ell$  describe the
 transformation properties of the local coordinates $z_a^\ell$ with respect to the $U(1)^2$ isometry of the manifold.  The parameters $a_\alpha^\ell$ describe the asymptotic values of the scalar field in the chart $\ell$ and of  the gauge fluxes
 \be\label{erre}
 r_\alpha^s = {1\over 2 \pi} \int_{D_s} F_\alpha={1\over 2 \pi} k_\alpha^\ell\int_{D_s}w_\ell
 \ee
 where $D_s$ are the divisors of $M$,  $s=1,\ldots b_2=\chi-2$, $b_2$ being the second Betti number.
 We see that the contribution of the gauge fluxes interlaces with that of point like instantons.

 We denote by $Z_{\rm full}$ the  contribution of $\mathfrak{F}_{\rm point}$ to the gauge partition function
 \be
Z_{\rm full}(a_\alpha,  r_\alpha^s, \epsilon_1, \epsilon_2,q,X)=\prod_{\ell=1}^\chi  Z_{\mathbb{C}^2} (a^\ell_\alpha ,\epsilon_1^\ell,\epsilon_2^\ell,q^\ell)
\ee
with
\be
a^\ell_\alpha = \mathfrak{a}_\alpha+k_\alpha^\ell \, \epsilon_1^\ell+ k_\alpha^{\ell+1} \, \epsilon_2^\ell
 =a_\alpha +r_\alpha^s \, C_s^{\ell } (\epsilon_1,\epsilon_2) \label{aul20}
 \ee
 where the  $ C_s^{\ell } (\epsilon_1,\epsilon_2) $ are some linear combinations of the $\epsilon$'s determined by the toric geometry (see below for details).  Finally $a_\alpha=a_\alpha^1 $ (or equivalently  $\mathfrak{a}_\alpha$), parametrizes the scalar vev in the first chart.
 On an open toric variety the final result would then be given just by the sum over the gauge theory fluxes of the above formula, producing a function of the asymptotic values of the scalar fields $a_\alpha$ at the boundary .
On a compact manifold one has instead to further proceed to integrate over these parameters, which are in this case normalizable zero modes of the dynamical fields in the path integral.

\paragraph{Integrating out the zero modes:}
\noindent
We proceed now to perform the path integral over the
zero modes of the fields. These zero modes are in the Cartan subalgebra
of the $SU(N)$ gauge group.  Since we are considering compact toric surfaces, which are simply connected, there are no zero modes for the one-form fields $A$ and $\Psi$. Moreover,
the zero modes of $B^+$ and $\chi^+$ are aligned along the Kahler form $\omega$
\bea\label{zeromodes}
\chi^+ &=&  \chi_\alpha\,h^\alpha\,\omega  \qquad , \qquad  B^+= b_\alpha\,h^\alpha\,\omega  \nn\\
\Phi &=& a_\alpha \, h^\alpha \qquad  ~~~, \qquad    \eta = \eta_\alpha h^\alpha
\eea
 The indices are raised and lowered using (\ref{cartannormaliztion}).   The zero modes are invariant configurations in the field space under the $V$-isometry action, therefore ${\mathcal Q}^2=0$ in this sector
 and the coefficients in \eqref{zeromodes} are constant.

 It is important to notice that the presence of bosonic and fermionic zero modes leads to an ambiguity in the definition of the path integral measure.
 Indeed the fermionic zero modes $\chi,\eta$ do not appear in the gauge theory action  (\ref{lagrangian}),
 moreover
 $Z_{\rm full}(a_\alpha)$ is meromorphic in $a_\alpha$ and if integrated over the zero modes $a_\alpha,\bar a_\alpha$ would yield a divergent result. To cure
 this ambiguity, we deform the action by adding the ${\cal Q}$-exact term
 \be\label{gf02}
{\cal Q} \mathcal{V}' = 2i s    {\cal Q} \,   \Tr[ {\rm Im}(\Phi) \, \chi ^+  \wedge \omega]\, ,
\ee
 where $s\in{\mathbb R}$ is a real gauge-fixing parameter \footnote{We take $s$ to be positive for the sake of simplicity. Actually the result of the zero modes integration does not depend on $s$.}.  Collecting the contributions from \eqref{GFFform} and (\ref{gf02})  one finds the zero modes action
 \be
S_{\rm zero\,modes}=
s \eta_\alpha\chi^\alpha + 2is\,{\rm Im}({a}_\alpha)b^\alpha
+\frac{g^2}{4}\, b_\alpha\, b^\alpha+2\pi i  \, b^\alpha \kappa_\alpha
\ee
where  we normalise the volume as $\int_M \omega\wedge\omega=1$ and we introduce the notation
\be\label{kappaalfa}
\kappa_\alpha={1\over 2\pi }\int_M\omega\wedge F_\alpha = \beta_\ell  \,  k_\alpha^\ell
\ee
with
\be
\beta_\ell={1\over 2 \pi} \int_M\omega\wedge w_\ell. \label{betaell}
\ee

\subsection{The $SU(2)$ case}

 Let us consider the $SU(2)$ case.
The integrals over $(\eta,\chi,b)$ can be easily performed
\bea
 \int   d\chi\, d\eta\,db
e^{\left[ \frac{s}{2}\, \eta\,\chi -\frac{g^2}{8} \, b^2 -2\pi{\rm i }\, b \, \kappa
-  {\rm i }s\,{\rm Im}(a)b \right]}
= \frac{s \sqrt{2\pi }}{g}  e^{-\frac{2}{g^2}
\left(s{\rm Im} (a) +2\pi\kappa \right)^2  }
\label{zzm}
\eea
  leading to
\be
Z_M=
{1\over 4 \pi^2} \sum_{\bf r}
  \frac{s\sqrt{2\pi}}{g}
\int_{{\mathbb C}}
da \wedge d\bar a \,
e^{-\frac{2}{g^2}
\left(s{\rm Im} (a) +2\pi\kappa \right)^2}
Z_{\rm full}(a,r^s,\epsilon_a,q,X)
\label{zero}\ee
 To perform the integral over $a$, we observe that the partition function $Z_{\rm full}(a,r^s,\epsilon_a,q,X)$ is a meromorphic function of the moduli with poles in the Cartan variable $a$ located along the real axis and at $\infty$.  Therefore, one can write
\be\label{lau}
Z_{\rm full}(a,r^s,\epsilon_a,q,X) =[{\rm regular \, part}]+
\sum_P
\frac{{\rm Res}_{P} Z_{\rm full}(a,r^s,\epsilon_a,q,X) } { a-P} +
\,\left[{\rm more\,\, singular\,\, terms}\right]
\ee
Since the theory is asymptotically free we expect that the regular part in $a$ can be regularized out of the integral. All the poles are crucially along the real axis.
  The contribution of the more singular terms vanishes upon integration over ${\rm Re} (a)$. The contribution of the second term leads to
\bea
  \int_{{\mathbb C}}
{ da\wedge d\bar a \over  4 \pi^2} \,
\frac{e^{-\frac{2}{g^2}
\left(s{\rm Im} (a) +2\pi\kappa \right)^2}}{a-P}
&=&
\int_{{\mathbb R}^2}\frac{dx\,dy}{2\pi^2 i } \frac{e^{-\frac{2}{g^2}
\left(sy +2\pi\kappa \right)^2}}{x+iy} = - \int_{{\mathbb R}} {dy \over 2 \pi}
\,e^{-\frac{2}{g^2}
\left(sy +2\pi\kappa \right)^2} {\rm sgn}(y) \nn\\
&=&  \frac{\sqrt{2} g}{2\pi s}\int_0^{\frac{2\sqrt{2}\pi\kappa}{g}}e^{-y'^2}dy'
=\frac{g}{ 2 s \sqrt{2\pi}}{\rm Erf}(\frac{2\sqrt{2}\pi\kappa}{g}) \label{intder}
\eea
where we wrote $a-P=x+iy$ and used
$\int_{{\mathbb R}}\frac{dx}{x+iy}=-(\pi i){\rm sgn}(y)$ and  ${\rm Erf}(x)=\frac{2}{\sqrt{\pi}}\int_0^x e^{-u^2}du$.  Plugging the result of the integral into (\ref{zero}) one finds
\be
Z_M= \ft12 \sum_{r^s} \sum_{ (k^1,k^2)}  {\rm Erf}\left({ 2\sqrt{2}\pi\kappa / g}   \right)  {\rm Res}_{a=a^{(k^1,k^2)}} Z_{\rm full}(a, r^s,\epsilon_a, q)\, .
\label{tie}
\ee
We notice that the whole dependence on $s$ cancels as expected.
We will see that the poles in the $a$ plane are labeled by two integers $(k_1,k_2)$ that can be grouped together in $r^s$.
In terms of these variables the partition function takes the form
\be
Z_M=
\ft12 \sum_{ k^\ell}
{\rm Erf}\left(
\kappa\sqrt{2\pi\tau_2}
\right)  {\rm Res}_{\mathfrak{a}=0} Z_{\rm full}(\mathfrak{a}, k^\ell,\epsilon_a, q)
\label{tie}
\ee
where, as usual, $\tau_2$ is the imaginary part of $\tau$.

\paragraph{Holomorphic decoupling limit:}

A decoupling limit of the gauge parameter can then be defined by tuning $g\to 0$ keeping $\tau$ finite
\footnote{We notice that we can tune the gauge -fixing parameter $g$ to any value independently from  $\tau$. Setting $g=0$ corresponds to the $\delta$-gauge.}.
In this limit, the {\rm Erf}-function reduces to a ${\rm sign} (\kappa)$ factor.
Alternatively,  the same result is obtained by performing first the limit $g\to 0$ and then the integral over the zero modes. In this way, the integral (\ref{zzm}) is replaced by
\bea
 \int   d\chi\,d\eta\,db
e^{ \left[ \frac{s}{2}\, \eta\,\chi  -2\pi{\rm i }\, b    \kappa -i  s \,{\rm Im}(a) b  \right]}
 =  \pi \, \delta \left( {\rm Im}(a) +\frac{2\pi\kappa}{s}   \right)
\label{zzm2}
\eea
The integral over $a$ reduces then to an integral over ${\rm Re}(a)$ along the line
${\rm Im}(a)=-2\pi\kappa /s$ and it can be  written as a sum over all the residues
weighted by the sign of $\kappa$
 \be\label{tien}
Z_M =\ft12 \sum_{ k^\ell } {\rm sign}(\kappa)   {\rm Res}_{\mathfrak{a}=0} Z_{\rm full}( \mathfrak{a}, k^\ell,\epsilon_a,q) =
\sum_{ k^\ell , \kappa>0 }    {\rm Res}_{\mathfrak{a}=0} Z_{\rm full}( \mathfrak{a}, k^\ell,\epsilon_a,q)  \, .
\ee
   We notice that the partition function depends on the choice of the Kahler form $\omega$,  only through the restriction $\kappa=\beta_\ell k^\ell>0$ in the sum over the $k$'s with $\beta_\ell$ the equivariant volumes defined in (\ref{betaell}). This implies in particular, that the partition function is piecewise constant along the space of Kahler forms with jumps across the walls where a Kahler cone chamber vanishes (see  Sect.\ref{section4} and Sect.\ref{section6} for details). 
   
   It is important to observe that the condition $\kappa>0$ in the sum can be viewed as a restriction to slope stable gauge bundles. Moreover, we will later show that only stable and semi-stable sheafs contribute to the sum (\ref{tien}), since the contributions coming from the
   tuplets $\{ k^\ell \}$ associated to unstable sheafs will always cancel against identical contributions with opposite signs coming from the
   tuplets in their Weyl orbit. This provides as highly non-trivial consistency check of the localization formula (\ref{tien}) and suggests   
     a generalization to higher rank that we will briefly sketch in the next section. 
  
 \subsection{The higher rank case}

A localization formula for the higher rank case can be obtained following {\it mutatis mutandis} the same steps, but now the derivation 
involve a multiple integration over the Cartan modes $a_\alpha$,  $\alpha=1,\ldots N-1$. Again the integration over the real parts of $a_\alpha$ picks the residues at $\mathfrak{a}_\alpha=0$ (let us say in the order $a_1>>a_2$) while that over the imaginary parts will now produce a generalized error function.
Again, one expects that in the limit $g\to 0$, this error function should reduce to a piece-wise constant sign-function with jumps in 
the moduli space of Kahler forms along the walls of marginal instability. A path integral derivation of this formula goes beyond the scope of this paper, but in analogy with the results for $SU(2)$ we expect again a formula given in terms of a sum of residues of $Z_{\rm full}$ over the tuplets $\{ k^\ell_\alpha \}$ associated to slope stable gauge bundles or equivariantly equivalent stable sheafs.  

  In this section, we review the concept of slope stability for  $U(N)$ gauge bundles. We refer the reader to Appendix D for a review on sheaf stability.
 
   A bundle $E$ is said to be (semi) stable if for every proper sub-bundle $G\subset E$ the slope 
 \be
  \mu(E)=\frac{\int_M c_1(E)\wedge\omega}{r(E)}
 \ee
  of the bundle is greater or equal than the slope $\mu(G)$ of the sub-bundle. Here $c_1(E)={1\over 2\pi } F(E) $ is the first Chern class 
  and $r(E)$ is the rank.  
The semi stability of the bundle is equivalent to the hermitian Yang-Mills equations via the Hitchin-Kobayashi correspondence \cite{Donaldson,Freed:1984xe
}. Namely, for an holomorphic vector bundle $E$, we have 
\be
\omega\wedge F_E=2\pi \mu(E) \, \omega\wedge\omega \, {\bf 1}_E\, .
\ee
Actually, if the vector bundle $E$ admits a sub-bundle $G$, then its gauge connection splits in blocks as\footnote{Notice that the specific block diagonal decomposition is obtained by fixing the Weyl group symmetry. }
\be
 A_E=\begin{pmatrix}    \star & v\\ v^\dagger & A_G  \end{pmatrix}
 \ee
and its curvature as
\be
 F_E=\begin{pmatrix} \star & \star \\ \star &   F_G + v\wedge v^\dagger    \end{pmatrix}.
 \ee
Inserting the last in the hermitian Yang-Mills condition above and projecting on the sub-bundle $G$ one finds that 
\be
\omega\wedge[F_G + v\wedge v^\dagger]=
2\pi \mu(E) \, \omega\wedge\omega \, {\bf 1}_G\, .
\ee
Taking the trace, integrating over $M$, and using that $\int \Tr (v\wedge v^\dagger) \wedge \omega  \geq 0$, we find  
the slope semi stability condition
\be
\mu(G) \leq  \mu(E)  \, \label{mumu}
 \ee
 Evaluating (\ref{mumu})  for all possible sub-bundles and using \eqref{kappaalfa}, we find the $N-1$ inequalities
\be
  \frac{1}{n}\sum_{\alpha=0}^{n-1}\kappa_{N-\alpha} \leq   \frac{1}{N}\sum_{\alpha=1}^{N-1}\kappa_{\alpha} 
\quad
{\rm for}
\quad n=1,\ldots N-1
  \label{acondition}
\ee
In particular for $N=2,3$ one finds  
 \bea
 SU(2): && \qquad \kappa \geq 0 \nn\\
  SU(3): && \qquad \kappa_1 +\kappa_2 \geq  0 \quad {\rm and } \quad  2\kappa_1 \geq \kappa_2     \label{slopecond}
 \eea
 As anticipated, the restriction $ \kappa>0$ in the localization formula sum (\ref{tien}), ensures that only slope semi (stable) bundes contribute to the partition function.  Similarly for $SU(3)$ we expect that the partition function should be given by a sum (with some signs determined by the path integral) restricted to bundles satisfying the slope stability condition (\ref{slopecond}).
 
 The equivariant version of slope stability conditions has been worked out by Klyacho in \cite{Klyachko1}. Equivariant stability is again defined by the requirement that the slope $\mu( F)$ of an equivariant subsheaf $F$ of the sheaf $E$ is smaller  than the slope of the sheaf itself $\mu(E)$. For SU(2) and SU(3) on $\mathbb{P}^2$ one finds (see Appendix for details)
 \bea \label{Kool0}
   SU(2): \quad && k^m \leq  k^n+ k^p           \nn\\
   SU(3): \quad   && k_1^m  \leq \ft13 \sum_{\alpha,\ell} k_\alpha^\ell           \quad , \quad   ~~~~~~~ k_2^m +k_2^n  \leq \ft13 \sum_{\alpha,\ell} k_\alpha^\ell   \nn\\
           && k_1^m +k_2^1+ k_2^2+k_2^3 \leq \ft23 \sum_{\alpha,\ell} k_\alpha^\ell     \quad , \quad   k_1^m +k_1^n+k_2^p \leq \ft23 \sum_{\alpha,\ell} k_\alpha^\ell  
\eea
  with $ m\neq n\neq p $ is always understood. There are 3 conditions for SU(2) and 12 for SU(3) where the two lines arise from one and two-dimensional sub-sheafs.  The two notions of stability are equivalent, so one expect that only bundles satisfying these more restrictive conditions will contribute to the partition function.

   In section  \ref{sectionsun}  we will apply these ideas to the $SU(3)$ theory and evaluate the first instanton corrections to the Donaldson partition function on $\mathbb{P}^2$.

 \subsection{Holomorphic anomaly equation}

The anomalous dependence in the gauge fixing parameter $g$ of the ${\cal N}=2$ partition function can be physically understood as coming from the contribution
of abelian anti-instantons. These are the fluxes of the gauge field  along the K\"ahler form, contributing to the zero-mode action as
it follows from \eqref{ttb}.
 The anomalous dependence can be obtained by taking the derivative of \eqref{tie} with respect to the coupling $g$.
By using the fact that
\be
\tau_{2}^{1/2}\frac{\partial}{\partial\bar\tau} {\rm Erf}(\kappa\sqrt{2\pi\tau_2}  )=
i\frac{\kappa}{\sqrt{2}} (q\bar q)^{\kappa^2/2}
\ee
one finds
\be\label{hae}
\tau_{2}^{1/2}\frac{\partial}{\partial\bar\tau}Z_{M}=\frac{\pi i}{2\sqrt{2}}
\sum_{ k^\ell }  \kappa (q\bar q)^{\kappa^2/2} {\rm Res}_{\mathfrak{a}=0} Z_{\rm full}( \mathfrak{a})
\ee
The $q^{\kappa^2\over 2}$ term cancels against a similar contribution coming from the classical part\footnote{ Remember that $\bar\tau\equiv\tau-i\frac{8\pi}{g^2}$ }.
This provides a generalization of the well known holomorphic anomaly equation in the ${\cal N}=4$ SYM discussed in \cite{Dabholkar:2020fde}.

An alternative derivation of the above equation for $SU(2)$ $\mathcal {N}=2$ SYM can also be given
by proceeding to the direct evaluation of the derivative of the partition function
with respect to the gauge coupling $g$. This leads to the calculation of the v.e.v. of a $\cQ$-exact operator
which is nonetheless different from zero due to boundary effects.
Indeed, by deriving with respect to $g$
one gets the following integral over the zero-modes
\be\label{deriv}
-\frac{4}{g} \frac{\partial}{\partial g} Z_M =  \int db\, \frac{d \mathfrak{a} \, d\bar{\mathfrak{a}} }{4\pi^2}\, d\chi\, d\eta \, {\cal Q} \,\left( \chi b\,  e^{-\frac{g^2}{8} b^2 - 2i\pi b \kappa}   Z_{\rm full} (\mathfrak{a}) \right)
\ee
where $Z_{\rm full}$ is a meromorphic function of $a$ (we omitted for simplicity its dependence on the other parameters which does not matter for the present argument).
By using the fact that on the zero modes $\cQ= \eta\frac{\partial}{\partial \bar{\mathfrak{a} }} + \chi\frac{\partial}{\partial b}$, we can absorb the fermionic zero modes $(\chi,\eta)$
and reduce \eqref{deriv} to a boundary term
\be
\int db\, \frac{d\mathfrak{a}\, d\bar{\mathfrak{a}}}{4\pi^2} \, \frac{\partial}{\partial \bar{\mathfrak{a} }} \left( b \, e^{-\frac{g^2}{8} b^2 - 2\pi i b \kappa} Z_{\rm full} (\mathfrak{a}) \right)
\ee
Now by using the analytic properties of $Z_{\rm full}(\mathfrak{a})$ and $\frac{\partial}{\partial \bar{\mathfrak{a}} }\frac{1}{\mathfrak{a}} = -\pi \delta^{(2)} (\mathfrak{a})$ we get
\be
-\frac{4}{g} \frac{\partial}{\partial g} Z_M =
{i \over 2} \sum_{ k^\ell } {\rm Res}  Z_{\rm full}  \int db\, b \, e^{-\frac{g^2}{8} b^2 - 2\pi i b \kappa} =
\sqrt{2} \tau_2^{3/2}
\sum_{k^\ell } \left({\rm Res}_{\mathfrak{a}=0    } Z_{\rm full} (\mathfrak{a})\right)  \kappa  \,
e^{-2\kappa^2\pi\tau_2}
\label{hono}
\ee
in agreement with \eqref{hae}. Alternatively one could evaluate the boundary term at infinity leading to the same result.
Let us observe that the derivation we outlined here is valid for general $\mathcal{N}=2$ gauge theories, the details of the theory depending on the explicit form of $Z_{\rm full}$.
The situation is very different from the $\mathcal{N}=4$ theory because of the
non renormalization of the gauge coupling and the corresponding vanishing
of the $U(1)_R$ anomaly. This is due to the
appearance of extra zero modes. In this case it is the non-abelian sector of the
$R$-symmetry group which is anomalous, inducing a non-trivial bundle structure of the zero-modes over the fixed point locus. The boundary term is therefore obtained in terms
of the non-trivial Chern classes of the $R$-symmetry bundle. This more intricate structure has been analysed in detail in \cite{Dabholkar:2020fde} from the viewpoint of the effective
abelian theory in the IR.

 \section{The partition function on $\mathbb{C}^2$ }
 \label{section3}
\setcounter{equation}{0}
 In order to set the notation, let us briefly recollect here the results for the partition function of the SYM on  $\mathbb{C}^2 $
 which are needed in the subsequent sections.
 For a pure $U(N)$  SYM on  $\mathbb{C}^2 $ we have a product of the classical, one loop and instanton contributions
 \be
 Z_{\mathbb{C}^2} =Z_{\rm classical,\mathbb{C}^2} \, Z_{\rm one-loop,\mathbb{C}^2} \, Z_{\rm inst,\mathbb{C}^2}
 \ee
  with $a$ parametrizing the expectation value at infinity of the scalar field in the vector multiplet, $\epsilon_a$ describing the $\Omega$-background and $q=e^{2\pi {\rm i} \tau}$ .  The classical  part is given by
     \begin{equation} \label{zclass0}
  Z_{\rm class,\mathbb{C}^2}  = q^{- { \sum_{u}  a_{u}^2   \over 2   \epsilon_1 \epsilon_2} }
  = q^{- {  \sum_{u,v}  a_{uv}^2  \over 4N   \epsilon_1 \epsilon_2} }
    \, ,
 \end{equation}
 with $a_{uv}=a_u-a_v$ with $u=1,\ldots,N$. The results for the $SU(N)$ theory are recovered by imposing the traceless condition $\sum_u a_u=0$.
The one-loop partition function reads
\bea
Z_{\rm 1loop,\mathbb{C}^2 }
 &=&  \prod_{u \neq v=1}^N    \frac{1}{\Gamma_2(a_{uv}) }
\label{zloop0}
 \eea
 with $\Gamma_2(x)$ the Barnes Gamma function  (see Appendix \ref{BDGF} for definitions and  details).
Finally the instanton partition function is given by a sum
  over the array $Y=\{ Y_u \}$ of  $N$ Young diagrams
 \be
    Z_{\rm inst,\mathbb{C}^2}
 = \sum_{Y}   q^{|Y|} \, Z_Y
 \label{zinst00}
\ee
where
  \bea
Z_Y   =    \prod_{u,v=1}^N  &&  \prod_{ (i,j)\in Y_u}
      \left[ a_{uv}+\epsilon_1(i- l_{vj})-\epsilon_2 \, (j-1-\tilde l_{ui})\right]^{-1} \nn\\
  \times && \prod_{ (i,j)\in Y_v}
      \left[a_{uv}-\epsilon_1(i-1- l_{uj})+\epsilon_2 \, (j-\tilde l_{vi})\right]^{-1}
        \label{zalbe}
\eea
      $\{ l_{uj} \} $  and $ \{ \tilde l_{ui} \}$  denote the length of the rows and columns respectively of the diagram $Y_u$, and  $|Y|$ counts the total number of boxes in $Y$. For instance in
 the case of the $SU(2)$ theory one finds, setting $a=a_{12}$
  \bea
  Z_{{\rm inst}, \mathbb{C}^2} (a ,\epsilon_1,\epsilon_2,q) &=& 1+
    {2\, q \over \epsilon_1 \, \epsilon_2\, (    (\epsilon_1+\epsilon_2)^2 - a^2 ) } \\
    &&+ q^2 { 2 ( 4\epsilon_1^2+ 4\epsilon_2^2-a^2) +17 \epsilon_1 \epsilon_2 \over \epsilon_1^2 \, \epsilon_2^2\, ( (\epsilon_1+\epsilon_2)^2- a^2 ) ( (\epsilon_1+2\epsilon_2)^2- a^2 )
     ( (2\epsilon_1+\epsilon_2)^2- a^2 )   }  +\ldots  \nn
    \eea
 More generally, the instanton partition function for $SU(2)$ can be written in the Zamolodchikov's form \cite{Poghossian:2009mk}
 \begin{equation} \label{zamol}
   Z_{\rm inst,\mathbb{C}^2}(a)=1+\sum_{m,n=1}^{
   \infty} \frac{q^{mn}R_{mn}Z_{inst}(m\epsilon_1-n\epsilon_2)}{(a+m \epsilon_1+n\epsilon_2)(-a+m \epsilon_1+n\epsilon_2)}
 \end{equation}
  \begin{equation}\label{rmn}
    R_{mn}=2\prod_{i=-m+1}^m \prod_{\substack{j=-n+1 \\ (i,j) \neq (0,0),(mn) }}^n \frac{1}{i \epsilon_1+j \epsilon_2}
 \end{equation}
 The instanton partition function can then be written in general as an infinite sum in $q^k$ with coefficients given by rational functions with poles at $a_u=\pm a_u^{(m,n)}$   where
  \be
  a_{uv}^{(m,n)} =m_{uv} \epsilon_1 + n_{uv} \epsilon_2
  \ee
  with $m_u,n_u$ some positive integers. On the other hand $Z_{\rm one-loop}$ (see (\ref{1looppositiva}) and (\ref{1loopnegativa}))
  has zeros exactly at these locations for $\epsilon_1 \epsilon_2>0$, so the full partition has no poles in this case.

\begin{remark}
 The partition function $Z_{\mathbb{C}^2}$ has poles if and only if $\epsilon_1 \epsilon_2<0$.
\end{remark}

\subsection{An abstruse duality}

In this section we show that the residues of the gauge partition function at the poles of its one-loop and instanton parts exactly coincide.

\subsubsection{The $SU(2)$ case}

We start by considering the $SU(2)$ case. We will show that the following identity holds
  \bea
   \label{signflip}
 \lim_{\mathfrak{a}\to 0}  { Z_{\mathbb{C}^2} (\mathfrak{a}+a^{(m,n)}  ) \over   Z_{\mathbb{C}^2} (\mathfrak{a}+\hat a^{(m,n)}  ) } = - \sign(\epsilon_1)
  \eea
with
\begin{equation}
    a^{(m,n)}=m \epsilon_1+n \epsilon_2, \qquad \hat{a}^{(m,n)}=m \epsilon_1-n \epsilon_2
\end{equation}
where $m,n$ are arbitrary non-zero integers. Similar formulae exists in the case in which we flip the sign of $m$  with $\sign(\epsilon_1)$ replaced by
$\sign(\epsilon_2)$.   To prove the duality relations we first observe that
\begin{equation}
   Z_{\rm{class},\mathbb{C}^2} ( \hat{a}^{(m,n)}) =   Z_{\rm{class},\mathbb{C}^2}( a^{(m,n)}) q^{mn}
   \label{zclass1}
\end{equation}
that follows from (\ref{zclass0}). On the other hand, using (\ref{zamol}) one finds that for $m n >0 $
 \begin{equation}
   {\rm Res}_{a=a^{(m,n)}} (Z_{\rm inst,\mathbb{C}^2} (a)) =  - {q^{mn} \, R_{mn} Z_{\rm inst,\mathbb{C}^2} (\hat{a}^{(m,n)})\over
   2(m \epsilon_1+n\epsilon_2)}
   \label{zinst1}
\end{equation}
 Finally, let us consider the one-loop partition function.
 To this aim, it is convenient to use the representation
 \be
\prod_i   \left( { x_i \over \Lambda } \right) ={\rm Exp}\left( -{d\over ds}\left[  {\Lambda^s\over \Gamma(s) } \int_0^\infty {dt\over t}  t^s\,  \sum_i  e^{-x_i t}  \right]_{s=0}  \right) \label{logrep0}
\ee
 to write products/determinants as sums/traces.  Using this representation one can write the one-loop partition function as
\be
Z_{\rm 1loop,\mathbb{C}^2 }  = {\rm Exp} \left(- {\diff \over \diff s}\left[  {\Lambda^s\over \Gamma(s) } \int_0^\infty {\diff t\over t}
{t^s\, \chi_{\mathbb{C}^2}(y,t_1,t_2) }\right]_{s=0}  \label{ge1e2} \right)
\ee
in terms of  the character  $\chi_{\mathbb{C}^2}$
  \bea \label{charc2}
\chi_{\mathbb{C}^2}(y,t_1,t_2) &=&      \frac{y+y^{-1} } {  (1-t_1)(1-t_2) }
\eea
  counting the Lie valued holomorphic funtions on $\mathbb{C}^2$  and
  \bea
   t_1 &=& e^{-t \, \epsilon_1} \,   , \qquad  t_2=e^{-t \, \epsilon_2 } \,  , \qquad     y=e^{-t \, a }.
  \eea
 The ratio between $Z_{\rm one-loop,\mathbb{C}^2}(\mathfrak{a}+a^{mn})$ and $Z_{\rm one-loop}(\mathfrak{a}+\hat{a}^{mn})$ can be computed in terms of the difference of the corresponding characters. For instance, for $\epsilon_1,\epsilon_2>0$, the difference of the associated characters in the limit  where $\mathfrak{y} =e^{-t\mathfrak{a}} \approx 1$ reduces to
\bea
  \chi_{\mathbb{C}^2}( \mathfrak y \, t_1^m t_2^n , t_1,t_2) - \chi_{\mathbb{C}^2}(  \mathfrak y \, t_1^m t_2^{-n} , t_1,t_2)  &=& \frac{\left(t_1^m t_2^n  \mathfrak y  + \mathfrak y ^{-1} t_1^{-m} t_2^{-n} - \mathfrak y   t_1^m t_2^{-n} - \mathfrak y ^{-1} t_1^{-m} t_2^{n}\right) }{(1- t_1)(1-t_2)}  \nn\\
  &\approx &   \sum_{i=-m}^{m-1}\sum_{\substack{j=-n \\ (i,j) \neq (0,0)}}^{n-1} e^{-t(i \epsilon_1 + j \epsilon_2)}  + \mathfrak y ^{-1}  \label{difchar}
\eea
One can recognize in the double sum of the right hand side of (\ref{difchar}) the character associated to the product (\ref{rmn})
\begin{equation}
    \frac{R_{mn}}{2(m \epsilon_1+n\epsilon_2)}=\prod_{i=-m}^{m-1} \prod_{\substack{j=-n \\ (i,j) \neq (0,0)}}^{n-1} \frac{1}{i \epsilon_1+j \epsilon_2}
\end{equation}
  We conclude then that
  \begin{equation}
    \frac{R_{mn}}{2(m \epsilon_1+n\epsilon_2)} \left[ \frac{Z_{\rm 1loop,\mathbb{C}^2}}{-\mathfrak{a} }\right]_{a=\mathfrak{a}+a^{(m,n)}} \approx - Z_{\rm 1loop,\mathbb{C}^2}|_{a=\hat{a}^{(m,n)}}
    \label{zone1}
\end{equation}
 Collecting (\ref{zclass1}), (\ref{zinst1}) and  (\ref{zone1}) one finds that the second line of (\ref{signflip}) is verified. A similar analysis  holds for $\epsilon_1<0$ leading again to (\ref{difchar}) with  $\mathfrak y ^{-1} $ replaced by   $\mathfrak y $ contributing an extra minus sign.

\subsubsection{The higher rank case}

  A generalization of the abstruse duality formula \eqref{signflip} holds for theories with unitary gauge groups of arbitrary rank. We have no proof of this duality but we checked its validity up to four instantons for $SU(3)$ and $SU(4)$ gauge theories. We claim that  the expansions of the gauge partition functions around the poles of its one-loop and instanton parts are related by
 \be
 \lim_{\mathfrak{a}_u\to 0} { Z_{\mathbb{C}^2} ( \mathfrak{a}_u+a_u^{(m,n)},\epsilon_a, q) \over Z_{\mathbb{C}^2} (\mathfrak{a}_u+\hat a_u^{(m,n)},\epsilon_a, q)}
 =  -\sign(\epsilon_1)
 \label{abduality2}\ee
 with
 \be
 a^{(m,n)}_u=m_u \epsilon_1+n_u \epsilon_2 \quad, \quad \hat a^{(m,n)}_u = a^{(m,n)}_u +n_1  \,\epsilon_2 (\delta_{uN} - \delta_{u1} )
 \ee
 and $(m_u,n_u)$ two ordered sequences of positive integers
 \be
 m_1>m_2>\ldots   m_N=0  \qquad ,\qquad n_1>n_2>\ldots >n_N=0
 \label{assump}
 \ee
 Explicitly
 \begin{equation}
 \begin{aligned}[c]
 a^{(m,n)}_1 &=m_1 \epsilon_1 + n_1 \epsilon_2 \\
 a^{(m,n)}_{\hat{u}} & =m_{{\hat{u}}} \epsilon_1 +n_{{\hat{u}}} \epsilon_2 \\
 a^{(m,n)}_N &=0
 \end{aligned}
 \qquad \qquad
 \begin{aligned}[c]
 \hat{a}^{(m,n)}_1 & =m_1 \epsilon_1   \\
 \hat{a}^{(m,n)}_{\hat{u}} & =m_{{\hat{u}}} \epsilon_1 +n_{{\hat{u}}} \epsilon_2 \\
 \hat{a}^{(m,n)}_N & =  n_1  \epsilon_2
 \end{aligned}
 \end{equation}
 with $\hat{u}$ running from $2$ to $N-1$.
 Similar identities hold for any choice of $SU(N)$ roots $\alpha_{v v'}=(\delta_{u v} - \delta_{u v'} )$.
 For the classical part one finds
 \begin{equation}
 Z_{\rm{class},\mathbb{C}^2} ( \hat{a}_u^{(m,n)}) =   Z_{\rm{class},\mathbb{C}^2}( a_u^{(m,n)}) q^{m_1 n_1 }
 \label{zclass2}
 \end{equation}
 that follows from (\ref{zclass0}).
 As before, the ratio of one-loop contributions  can be extracted from the difference of the one-loop characters
 \bea  \label{diffchar}
 \Delta\chi(t_1,t_2)  =    \chi_{\mathbb{C}^2} ( \mathfrak y_u\,   y_u^{(m,n)},t_1,t_2) -  \chi_{\mathbb{C}^2} (  \mathfrak y_u\, \hat y_u^{(m,n)}  ,t_1,t_2 )
 \eea
 with  $y_u^{(m,n)}=e^{-t \,a_u^{(m,n)} }$ , $\hat y_u^{(m,n)}=e^{-t\, \hat a_u^{(m,n)} }$,  $\mathfrak y_u=e^{-t \mathfrak{a}_u} $.
 Plugging these formulae into (\ref{diffchar}) one finds generically  for $\epsilon_1,\epsilon_2>0$  in the limit $\mathfrak{y}_u \approx 1$
 \bea
 &&  \Delta\chi (t_1,t_2)   = {1  \over (1-  t_1)  (1-  t_2 ) }    \sum_{u,v}    \mathfrak{y}_{uv} t_1^{m_{uv} }    t_2^{n_{uv} }      \left( 1-
 t_2^{n_1 (\delta_{uN}+\delta_{v1} -\delta_{u1}-\delta_{vN} ) }  \right)   \nn\\
 && \approx    { 1   \over (1-  t_1)  (1-  t_2 ) }   \left[  (t_1^{m_1} -t_1^{-m_{1} } )  (t_2^{n_1} -t_2^{-n_{1}})
 +\sum_{ \hat u=2}^{N-1}  (   t_1^{m_{ \hat u} } -t_1^{-m_{1 \hat{u}}})  (   t_2^{n_{ \hat u} } -t_2^{-n_{1 \hat{u}}}) \right. \nn\\
 && \qquad \qquad +  \left.   \sum_{ \hat u=2}^{N-1}  (   t_1^{m_{1 \hat u} } -t_1^{-m_{\hat{u}}})  (   t_2^{n_{ 1\hat u} } -t_2^{-n_{ \hat{u}}})   \right] +
 \sum_{ u=1}^{N-1} ( \mathfrak{y}_{uN}^{-1} + \mathfrak{y}_{1u}^{-1}    -2  )    \label{difchar2} \\
 && \approx   \sum_{i=-m_1}^{m_{1}-1}  \sum_{j=-n_1}^{n_{1}-1} \,  t^{i}_1 \, t^{j}_2  + \sum_{ \hat u=2}^{N-1}
 \left[  \sum_{i=-m_{1 \hat{u}} }^{m_{\hat u } -1} \sum_{j=-n_{1\hat{u}}}^{n_{\hat u} -1}  \,  t^{i}_1 \, t^{j}_2+
 \sum_{i=-m_{\hat{u}} }^{m_{1\hat u }-1 }     \sum_{j= -n_{\hat{u}}}^{ n_{1\hat u}-1} \,  t^{i}_1 \, t^{j}_2 \right] +
 \sum_{ u=1}^{N-1} ( \mathfrak{y}_{uN}^{-1} + \mathfrak{y}_{1u}^{-1}    -2  ) \nn
 \eea
 leading to
 \begin{eqnarray} \label{PolP0}
 \frac{Z_{\rm 1loop,\mathbb{C}^2}( \mathfrak{a}_u+a_u^{(m,n)} ) } {Z_{\rm 1loop,\mathbb{C}^2}({\mathfrak{a}_u+\hat a_u^{(m,n)}}) }
 \underset{\mathfrak{a}_u\to 0}{\approx}  - P(\epsilon_1,\epsilon_2)\,  \mathfrak{a}_1 \prod_{\hat{u}=2}^{N-1} ( \mathfrak{a}_{\hat u}  \mathfrak{a}_{1\hat u}  )
 \end{eqnarray}
 with
 \bea
 P(\epsilon_1,\epsilon_2) &=&
 \prod_{i=-m_1}^{m_{1}-1}  \prod_{j=-n_1}^{n_{1}-1} \, (i \epsilon_1+j \epsilon_2)  \prod_{ \hat u=2}^{N-1}
 \left[  \prod_{i=-m_{1 \hat{u}} }^{m_{\hat u } -1} \prod_{j=-n_{1\hat{u}}}^{n_{\hat u}-1 }  \, (i \epsilon_1+j \epsilon_2)
 \prod_{i=-m_{\hat{u}} }^{m_{1\hat u }-1 }     \prod_{j= -n_{\hat{u}}}^{ n_{1\hat u}-1} \, (i \epsilon_1+j \epsilon_2) \right]\nn\\
 \eea
 
 In the case where $m_{u1} n_{u1} = (m_u-m_1)(n_u-n_1) = 0$ for a given u, the corresponding term  $\mathfrak{a}_{1\hat u}  $ in (\ref{PolP0}) is missing.
 For the ratio of the instanton contributions one finds
 \begin{eqnarray} \label{PolP0inst}
 \frac{Z_{\rm inst,\mathbb{C}^2}( \mathfrak{a}_u+a_u^{(m,n)} ) } {Z_{\rm inst,\mathbb{C}^2}({\mathfrak{a}_u+\hat a_u^{(m,n)}}) }
 \underset{\mathfrak{a}_u\to 0}{\approx}  {q^{m_1 n_1} \over P(\epsilon_1,\epsilon_2)\,  \mathfrak{a}_1 \prod_{\hat{u}=2}^{N-1} ( \mathfrak{a}_{\hat u}  \mathfrak{a}_{1\hat u}  )}
 \end{eqnarray}
 Collecting  (\ref{zclass2}),(\ref{PolP0}) and (\ref{PolP0inst})  one finds
 \be
 Z_{\mathbb{C}^2} (a_u^{(m,n)},\epsilon_a, q) \approx    -   Z_{\mathbb{C}^2} (\hat a_u^{(m,n)},\epsilon_a,q)
 \ee
 as claimed.  For $\epsilon_1<0$ the same  results are found with a flipped overall sign.

 For example for $SU(3)$ taking $m_1=m_2=n_1=n_2=1$, one finds from (\ref{PolP0})
 \begin{eqnarray} \label{PolP}
 \frac{Z_{\rm 1loop,\mathbb{C}^2}( \mathfrak{a}_u+a_u^{(m,n)} ) } {Z_{\rm 1loop,\mathbb{C}^2}({\mathfrak{a}_u+\hat a_u^{(m,n)}}) }
 \underset{\mathfrak{a}_u\to 0}{\approx}  - \mathfrak{a}_1\mathfrak{a}_2  \epsilon_1 \epsilon_2 (\epsilon_1+\epsilon_2)^2
 \end{eqnarray}
 %
 %
 %
 and
 \bea
 Z_{\rm inst,\mathbb{C}^2} &=&1+ 2 q {   a_{1}^2+a_{2}^2-a_{1} \, a_{2} -3 ( \epsilon_1+\epsilon_2)^2    \over \epsilon_1 \, \epsilon_2\, (    a_{1}^2 - (\epsilon_1+\epsilon_2)^2  )
 	(a_{2}^2 -   (\epsilon_1+\epsilon_2)^2  ) ( a_{12}^2-  (\epsilon_1+\epsilon_2)^2   )    } +\ldots \nn\\
 & \approx &  {    q\,Z_{\rm inst,\mathbb{C}^2} (\hat a_u^{(m,n)},q)  \over \epsilon_1 \, \epsilon_2\,  (\epsilon_1+\epsilon_2)^2   \mathfrak{a}_{1} \mathfrak{a}_{2}           }    +\ldots  \label{zinstaah}
 \eea

 \section{The partition function on $M$}
 \label{section4}
\setcounter{equation}{0}
 The gauge theory partition function on a compact toric $M$ \eqref{tien} is given as a residue formula of the exact semiclassical integrand $Z_{\rm full} $. This in turn
 can be written as a product of a classical, one-loop and instanton contribution
 \be
Z_{\rm full} (a_u,k_u^\ell)  =
\prod_{u \neq v}^N Z_{\rm classical}(a_{uv}, k^\ell_{uv} ) \, Z_{\rm one-loop} (a_{uv}, k^\ell_{uv})\, Z_{\rm inst}(a_{uv}, k^\ell_{uv})
\ee
 each one given as a product of contributions for each toric patch. In the case of $SU(2)$, one can simply relabel  the variables
 $(a_{12},k^\ell_{12}) \to (a,k^\ell)$. The formulae for $Z_{\rm one-loop} (a_{uv}, k^\ell_{uv})$, $\, Z_{\rm inst}(a_{uv}, k^\ell_{uv})$ $SU(N)$ follow from those of $SU(2)$ by sending  $ (a,k^\ell)$ to
  $(a_{uv},k^\ell_{uv}) $ and taking the product over all pairs $(u,v)$ with $u\neq v$. Without loss of information we can then
  restrict to the $SU(2)$ case. The classical part is given explicitly for the $SU(N)$ case.  In the following, after fixing the relevant data of the toric geometry,  we compute the classical, one-loop and instanton contributions separately.

 \subsection{The toric data}

 As explained in section \ref{section2}, the function $Z_{\rm full}(a, k^s,\epsilon_a,q)$  can be written as a product of
 partition functions $Z_{\mathbb{C}^2} (a^\ell,\epsilon_1^\ell,\epsilon_2^\ell,q^\ell)$ accounting for the contributions
 of instantons localized at the origins of each chart $\ell=1,\ldots \chi$ covering the manifold $M$.
 The equivariant parameters $\epsilon_a^\ell$  describe the
 transformation properties of the local coordinates $z_a^\ell$ with respect to the action of the vector field $V$ on the manifold.
 For a toric manifold, they can be determined  recursively (see Appendix \ref{toricapp} for details)
 starting from $( \epsilon_1^1,\epsilon_2^1) = ( \epsilon_1,\epsilon_2)$ via the  relation
\bea
 (\epsilon_1^{\ell+1},\epsilon_2^{\ell+1} ) &=& (- C_{\ell+1, \ell+1}\, \epsilon_1^\ell+\epsilon_2^\ell,-\epsilon_1^\ell)   \label{rece0}
 \eea
with  $C_{\ell \ell'}=D_\ell \cdot D_{\ell'}$ the intersection matrix  between
 the equivariant  divisors $D_\ell$.  We recall that only $b_2=\chi-2$ divisors are homotopically  independent,
so we can take the subset $\{ D_s \}$, $s=1,\ldots b_2$  as a basis of homotopically independent cycles.

 Similarly, the scalar vevs, which parametrise the solutions of (\ref{equi}), can be written as in (\ref{aul20}). Finally, as anticipated in Sect.\ref{section2}, the Donaldson observables are characterised by the choice of an equivariant polyform (\ref{donaldsonsource}).
 We normalise the two forms
   $\{ w_\ell \}$ such that
  \be
\frac{1}{(2 \pi)^2}\int_M\, w_\ell \wedge w_\ell'={1\over 2\pi} \int_{D_\ell}  w_{\ell'}= C_{\ell \ell'}
\ee
 As we said already, in presence of Donaldson observables, the induced gauge coupling in each chart is given by $q^\ell = q \, e^{\Omega_\ell}$ with ${\Omega_\ell}$ given by the zero form part of ${\cal T}$ evaluated at the fixed point, $z_1^\ell=z_2^\ell=0$,
in chart $\ell$.

\paragraph{Classical term:}
 The contribution of the classical action to the partition function on $M$ can be written as 
 \be
Z_{\rm class}( \mathfrak{a}, {\bf k}, \epsilon_1, \epsilon_2,q,X)=\prod_{\ell=1}^\chi  Z_{{\rm class},\mathbb{C}^2} (a^\ell,\epsilon_1^\ell,\epsilon_2^\ell,q^\ell)=q^{\Delta_k+c_1^2 \frac{N-1}{2N} }\, e^{\Delta^\Omega}
\ee
with $c_1^2= \int_M c_1 \wedge c_1 = \frac{1}{(2 \pi)^2}\int_M {\rm tr} F \wedge {\rm tr} F$  and
 \bea
     \Delta_{\bf k} &=&    - \sum_{u,v} \sum_{\ell=1}^\chi    {\left( \mathfrak{a}_{uv}+ k^{\ell}_{uv} \epsilon_1^\ell+ k^{\ell+1}_{uv} \epsilon_2^\ell \right)^2 \over 4 N  \epsilon^\ell_1 \epsilon^\ell_2}
    =-\sum_{u,v}{C_{\ell m} k_{uv}^\ell k_{uv}^m\over 4 N}\nn\\
      \Delta^{\Omega} &=& - \sum_{u,v} \sum_{\ell=1}^{\chi}     { \left( \mathfrak{a}_{uv}+ k_{uv}^{\ell} \epsilon_1^\ell+ k_{uv}^{\ell+1} \epsilon_2^\ell \right)^2  \Omega_\ell \over 4 N  \epsilon_1^{\ell} \epsilon_2^{\ell} } \label{ids}
    \eea
    We notice that the partition function does not depend on an overall U(1) shift of the fluxes $k_u^\ell \to k_u^\ell +c^\ell$, so we can use
      this freedom for example to set $c_1^2=0,1,..[N/2]$ in the case of $SU(N)$ theory.

    In deriving the right hand side of  (\ref{ids}) we used the identities
   \bea
    \sum_{\ell=1}^\chi  {1 \over     \epsilon^\ell_1 \epsilon^\ell_2}  &=&
       \left( {1 \over     \epsilon^{\ell+1}_2 }+{1\over \epsilon^\ell_1} \right) =0 \nn\\
        \left( {\epsilon_1^{\ell} \over     \epsilon^{\ell}_2 }+{\epsilon_2^{\ell-1}\over \epsilon^{\ell-1}_1} \right) &=& C_{\ell \ell}
    \eea
 following from (\ref{rece}). Remarkably, the classical partition function on $M$  without observables does not depend on the scalar vev $\mathfrak{a}$.

 \paragraph{One loop term:}
 By making use of (\ref{zloop0}), the one loop  partition function on $M$
 can be written as the product
  \be
Z_{\rm one-loop}( \mathfrak{a}, {\bf k}, \epsilon_1, \epsilon_2,q,X)=\prod_{\ell=1}^\chi  Z_{{\rm one-loop}, \mathbb{C}^2} (a^\ell ,\epsilon_1^\ell,\epsilon_2^\ell,q^\ell)
\ee
 We will first prove that, although each factor is given by an infinite product, the total result involves only a finite number
 of factors. To see this, it is convenient to consider the character representation of the one-loop contributions, so from (\ref{charc2}) we get
   \be
\chi_{\rm M}({\bf k} | y,t_1,t_2 )=   \sum_{\ell=1}^\chi {   y  \left(t_1^\ell \right)^{k^{\ell} } \, \left(t_2^\ell \right)^{k^{\ell+1}}
+ y^{-1}  \left(t_1^\ell \right)^{-k^{\ell} } \, \left(t_2^\ell \right)^{-k^{\ell+1}}
\over (1- t_1^\ell)(1-t_2^\ell)   } = \sum_{m,n,p} d^{\bf k}_{m n p } \, t_1^m t_2^n  \, y^p
\label{charonetty}
\ee
 It is easy to see that  $\chi_{M}({\bf k} | y, t_1,t_2 )$ is a finite polynomial given that it is a rational function and that it has no poles in the limit where  one of the $t_a^\ell$ goes to one.  For instance, for a given $\ell$, in the limit where $t_1^\ell\approx 1$, the two terms $\ell$ and $\ell+1$ in the sum lead to
    \be
\chi_{\rm oneloop}\big|_{t_1^\ell\approx 1}  \approx    y \left[
 {\left(t_1^\ell \right)^{k^{\ell} } \, \left(t_2^\ell \right)^{k^{\ell+1}}  \over (1- t_1^\ell)(1-t_2^\ell)}
 +{ \left(t_1^{\ell+1} \right)^{k^{\ell+1} } \, \left(t_2^{\ell+1} \right)^{k^{\ell+2}}  \over (1- t_1^{\ell+1})(1-t_2^{\ell+1})   }
 \right]+\ldots   \approx   0+{\rm regular~terms}
\label{charonetty1}
\ee
  where we used that $(t_1^{\ell+1},t_2^{\ell+1})=(t_2^\ell (t_1^{\ell})^{C_{\ell \ell}}, (t_1^\ell)^{-1})$ as it follows from (\ref{rece0}). Consequently, the one-loop partition function
 can be written as the finite product
 \bea
Z_{\rm oneloop}(\mathfrak{a},{\bf k},\epsilon_1,\epsilon_2)  =\prod_{m,n,p}    ( p\, a +m \epsilon_1+n \epsilon_2)^{-d_{mn p}^{\bf k} }
\label{zonetty1}
\eea
  with $p=\pm$ and where $d^{\bf k}_{mnp }$ are the expansion coefficients of the one-loop character.

 \paragraph{Instanton term:}
 The instanton contribution is
 \be
Z_{\rm inst}(\mathfrak{a}, {\bf k}, \epsilon_1, \epsilon_2,q,X)=\prod_{\ell=1}^\chi  Z_{{\rm inst}, \mathbb{C}^2} (a^\ell ,\epsilon_1^\ell,\epsilon_2^\ell,q^\ell)
\ee
 with $Z_{{\rm inst}, \mathbb{C}^2}$ given as a sum over Young diagrams contributions \eqref{zinst00}.

\subsection{ The residue sum}

 In this section we study the residues of $Z_{\rm full}$ in the $SU(2)$ case.  We start by focusing on the contribution of a single chart
 $Z_{\mathbb{C}^2}(a^\ell,\epsilon_a^\ell,q^\ell)$.
 Near $a\approx 0$ this function can have a zero or a pole depending on the relative signs of the $\epsilon^\ell$'s and whether or not the $k^\ell$,$k^{\ell+1}$s are zero. The different cases are listed in the following:
\begin{itemize}
\item
when $k^\ell, k^{\ell+1} \neq 0$, the function $Z_{\mathbb{C}^2}(a^\ell,\epsilon_a^\ell,q^\ell)$ has a pole if  $\epsilon_1^{\ell}\epsilon_2^{\ell}<0$ and  is regular otherwise
\item when $k^\ell=0$, the result is the same if $\epsilon_2^{\ell}<0$ and gets suppressed by an extra factor of $a$ if $\epsilon^{\ell}_2>0$.
Similar suppression factors are obtained for $k^{\ell+1}=0$ in the case of
 $\epsilon_1^{\ell}<0$ and $\epsilon_1^{\ell}>0$ respectively.
 \end{itemize}
Using the fact that in a compact toric manifold there are two and only two patches  with $\epsilon_1^\ell \epsilon_2^\ell >0$ (see
  Remark \ref{numberofregs} in Appendix), we conclude that

  \begin{remark} \label{chim2}

  The pole of $Z_{\rm full}$ associated to a tuplet of non-vanishing integers $\{ k^\ell \}$ is always of order  $\chi-2$.
  On the other hand, since $\epsilon_1^{\ell} \epsilon_2^{\ell+1}$ is always negative, every vanishing $k^\ell$ in the tuplet reduces the order of the pole by one.

\end{remark}

  Now let us consider the sum over the residues of  $Z_{\rm full}$. It is convenient to introduce the following
 operator
  \be
   {\cal P}_{\ell}  (k^{\ell'}) = (-1)^{\delta_{\ell\ell'}} k^{\ell'}
  \label{kset}\ee
  that flips the sign of the $\ell^{\rm th}$ component of the vector ${\bf k}$.

       Then we state the following property
   \be \label{idzero2}
 {\rm Res}_{a=0} \,  \prod_{\ell=m}^{m+\chi-2}   (1+{\cal P}_\ell)\, Z_{\rm full}(\mathfrak{a}, {\bf k},\epsilon_a,q) = 0  \,\,\, \forall m \in \{1, \ldots, \chi\}
\ee
   To prove this, we recall that for any $k$, the full partition function has at most $\chi-2$ poles. Let us consider first the $\chi=3$ case.
   In this case one has a single pole so each term in the product contributes with its residue. According to the abstruse duality \eqref{signflip},
   under a flip of the sign $k^\ell$, this term picks up the sign $\epsilon_1^{\ell} \epsilon_2^{\ell+1}=-1$, so the two contributions cancel against each other.
   Similarly for $\chi=4$ one can achieve a similar cancellation after flipping two signs in (\ref{kset}).
   A detailed derivation of the general case is presented in the Appendix \ref{proof}.

   An important consequence of (\ref{idzero2})  is that the sum over all tuplets $\{ k^\ell \}$ of the residues of $Z_{\rm full}$ cancels, so that a non-trivial result is found for the
   sum (\ref{tien}) weighted  by $\sign(\kappa)$.

   It is important to observe that the residue of $Z_{\rm full}$ for a given ${\bf k}$ gives in general a result involving negative powers of $q$ and
  divergent contributions in the non-equivariant limit $\epsilon_a\to 0$, while the partial sum
  \be \label{zeta}
   Z_{\rm orbit}( {\bf k},\epsilon_a,q)  = \frac{1}{2^{\zeta_{\bf k}} } {\rm Res}_{\mathfrak{a}=0}\left[    \prod_{\ell=1}^{\chi}   (1+{\cal P}_\ell) \sign(\kappa) Z_{\rm full}(\mathfrak{a}, {\bf k},\epsilon_a,q) \right]
  \ee
  involving the residue of $Z_{\rm full}$ and all its sign flips does not include the terms with negative powers of $q$, although it still can be divergent in the non-equivariant limit.
  Here $\zeta_{\bf k}$ is the number of zero entries in $\{ k^\ell \}$. This allows us to write the partition function as a sum
  over orbits labeled by tuplets  $\{ k^\ell \}$ of non-negative integers
   \be\label{zetaorbit}
   Z_M(\epsilon_a,q)=\sum_{{\bf k} \geq 0} Z_{\rm orbit}( {\bf k},\epsilon_a,q)
     \ee

Moreover, the orbits of the vector $\{ k^\ell \}$ with positive components only can be classified into three groups
\bea
{\rm Stable} && \forall \, k^i   \qquad    2 \beta_i \, k^i  < \sum_{\ell=1}^\chi \beta_\ell \, k^\ell \label{st} \\
{\rm Semi~stable} && \exists \, k^i   \qquad   2 \beta_i \, k^i  = \sum_{\ell=1}^\chi \beta_\ell \, k^\ell \label{semist} \\
 {\rm Unstable} && \exists \, k^i   \qquad   2 \beta_i \, k^i  > \sum_{\ell=1}^\chi \beta_\ell \, k^\ell   \label{unst}
\eea
 It is easy to see that the contributions from unstable orbits exactly cancel.
 In this case one can write
  \bea
   Z_{\rm unst\,orbit}( {\bf k},\epsilon_a,q)  &=& \frac{1}{2^{\zeta_{\bf k}} } {\rm Res}_{\mathfrak{a}=0}\left[ (1+{\cal P}_i)    \prod_{\ell\neq i}^{\chi}   (1+{\cal P}_\ell) \sign(\kappa)  Z_{\rm full}(\mathfrak{a}, {\bf k},\epsilon_a,q) \right]\\
 &=&  \frac{1}{2^{\zeta_{\bf k}} }  {\rm Res}_{\mathfrak{a}=0}  \left[ (1+{\cal P}_i)   \sign(\kappa)  \prod_{\ell\neq i}^{\chi}   (1+{\cal P}_\ell)  Z_{\rm full}(\mathfrak{a}, {\bf k},\epsilon_a,q)\right] =0\nn
  \eea
where in the last line we used the fact that if ${\bf k}$ belongs to an unstable orbit
 all the flips in signs of the $k^{\ell\neq i}$  do not change the sign of $\kappa$. The last equation follows then from (\ref{idzero2}).

Finally in the case of semi-stable orbits, some contributions are missing since $\kappa=0$ for specific choices of signs. 
The contribution of the tuplet $\{k^\ell\}$ from a semi stable orbit is two times less than that of the same tuplet for a stable orbit.
\par One can see a correspondence between the classification of tuplets of fluxes $\{ k^\ell \}$ (\ref{st}-\ref{unst}) and Klyachko's classification of sheaves on a manifold (and therefore between the poles of the partition function and the sheaves on a manifold)\cite{Klyachko1,Klyachko2}. According to Klyachko every sheaf is characterized by $\chi$ filtrations of vector spaces. If we identify the fluxes $\{ k^\ell \}$ with the positions of the jumps in the filtrations, we will see that (\ref{unst}) corresponds to the filtrations defining only unstable sheaves, while (\ref{st}) is the necessary condition to have a stable sheaf among all the sheaves described by corresponding filtrations. The intermediate condition (\ref{semist}) guarantees that there is a semistable sheaf among all of the sheaves described by the corresponding filtrations (see Appendix \ref{Klyachko} for the details). The fact that the contribution of the unstable orbits vanish is in agreement with the known fact that unstable sheaves do not contribute to the Donaldson invariants. The correspondence between the poles of the partition function and the sheaves on a toric manifold was first noted in \cite{Bershtein:2015xfa}.
\par An interesting observation is that the contribution of an orbit $Z_{\rm orbit}( {\bf k},\epsilon_a,q) $ changes exactly when some of the sheaves defined by the corresponding filtrations change their stability type.

\subsection{Comparison against wall-crossing formulae}

In this section we check that the  localization formula for the Donaldson partition function  agrees with the results obtained via wall-crossing. We focus on $SU(2)$ SYM on a compact toric manifold with $\chi>3$.
Let us consider the difference between the partition functions for two different choices of the Kahler form, $\omega$ and $\omega'$.
The localization formula for the difference can be written in the form
\begin{equation} \label{difdon}
   Z^{\omega}_{M}-Z^{\omega'}_{M}=  2  \sum_{  \substack{ {\bf r} \in \mathbb{Z}^{2(\chi-2)}+c_1 \\ \kappa \kappa' < 0 } }  {\rm Res}_{a=\infty}
Z_{\rm full}(a^1, {\bf r}, \epsilon_1, \epsilon_2,q,X)
\end{equation}
 for gauge bundles of fixed torsion class $c_1=0,1$.   The sum with  $\kappa \kappa'<0$ can be alternatively written as twice the sum over  those $r^s$ satisfying  $\kappa>0$ and $\kappa'<0$.
 This sum can be interpreted as the contribution of the jumps made by the partition function when crossing the walls  $\kappa=0$ in the space of K\"ahler forms.
 \par We denote by $\hat{r} =k^\ell \omega_\ell|_{\rm 2-form} = [F] \in H^2(M)$ the non-equivariant class of the gauge field strength $F$.
 The two-form $\hat{r}$ is determined by the coefficients $r^s$. We notice that the condition $\kappa \kappa' \sim ( \int \omega \wedge \hat{r})( \int\hat{r} \omega' \wedge )< 0$ requires that $\hat{r}$ is a space-like form
  $\int \hat{r} \wedge \hat{r} <0$ since  $\omega$, $\omega'$ are time-like forms $\int \omega \wedge \omega >0$, $\int \omega' \wedge \omega' >0$\footnote{We remind that since $b_2^+=1$ the space $H^2(M)$ with scalar product $\int a \wedge b $ has a Minkowski-like signature} and belong to the K\"ahler cone.
Consequently
\begin{equation}
\Delta_{\textbf{k}}=-\frac{1}{8}\int \hat{r}\wedge \hat{r} >0
\end{equation}
 and therefore the difference (\ref{difdon}) has a well defined weak coupling expansion given by truncating the sum over the $r^s$
 to a given value of $\Delta_{\textbf{k}}$. On the other hand for manifolds with $\chi=3$, namely $\mathbb{P}^2$, $\kappa \kappa'=r^2>0$ so the difference vanishes and no walls are found as expected.

  In the mathematical language, following \cite{Gottsche:2006tn}, a wall is defined as follows.

\par \textbf{Definition.} For every class $\xi \in H^2(M,\mathbb{Z})\setminus \{0\}$ such that $\langle \xi \cdot \xi \rangle < 0$ there is a wall $W^{\xi}:=\{ L \in H^2(M,\mathbb{R}) \, | \,  \langle L \cdot \xi \rangle =0 \}$ of type $c_1, \, c_2$  if
\par 1. $W^{\xi} \neq \emptyset$;
\par 2. $\xi + c_1$ is divisible by 2 in $H^2(M,\mathbb{Z})$;
\par 3. $c_1^2-4c_2 \leq \langle \xi \cdot \xi \rangle $.
\bigskip
\par As it is shown in \cite{Gottsche:2006tn} the difference between the equivariant Donaldson invariants $\tilde{\Phi}$ in a chamber containing a polarization $\omega$ and a chamber containing a polarization $\omega'$ can be found as
\begin{equation} \label{wallcross}
\tilde{\Phi}_{c_1}^{\omega}-\tilde{\Phi}_{c_1}^{\omega'}=\sum_{\xi} \tilde{\delta}_{\xi}
\end{equation}
where the sum goes over all classes $\xi$ defining the walls of given $c_1$ and arbitrary $c_2$ such that $\int \omega' \cdot \xi < 0 < \int \omega \cdot \xi $ and the  contribution from a class $\xi$ is given by
\begin{equation} \label{deltaxi}
\tilde{\delta}_{\xi} = \res_{a=\infty} \prod_{l} Z_{\mathbb{C}^2}(\epsilon_1^{(l)}, \, \epsilon_2^{(l)}, \, a - i^*_{P_l} \tilde{\xi} , \, q \, e^{i^*_{P_l}\Omega})
\end{equation}
with $\Omega$ being a form defining the observable and $\tilde{\xi}$ being any equivariant extension of the class $\xi$.
We identify the class $\xi$ with the non-equivariant two-form $\hat{r}$. Then taking into account that $a^\ell-a^{\ell-1}=i^*_{P_{\ell-1}}\tilde{\xi} - i^*_{P_\ell} \tilde{\xi}$
(see the localization theorem in \cite{Goresky}) we see that
formulae
(\ref{difdon}) and (\ref{wallcross}) match up to an overall numerical coefficient in the definition of the Donaldson invariants.

\section{ Donaldson invariants for $\mathbb{P}^2$}
\label{section5}
\setcounter{equation}{0}
In this section we consider $SU(2)$ SYM on $\mathbb{P}^2$.
The results  on $\mathbb{P}^2$ are well known for any choice of $c_1$, so one can use them to test our approach.

\subsection{Geometric data}

 \begin{figure}[!h]
\centering
\[
 \begin{array}{ccc}
\begin{tikzpicture}[scale=0.7]
\draw [->] (0,0) -- (0,2) node[above] {$v_2$};
\draw [->] (0,0) -- (2,0) node[right] {$v_1$};
\draw [->] (0,0) -- (-2,-2) node[left] {$v_3$};
\node at (1,1) {$\sigma_1$};
\node at (1,-1) {$\sigma_2$};
\node at (-1.5,0) {$\sigma_3$};
\end{tikzpicture}
& ~~~~~~~~~~~~~~&
\begin{tikzpicture}[scale=0.7]
\draw [->] (0,0) -- (2,0) node[right] {$v^*_2$};
\draw [->] (0,0) -- (-2,0) node[left] {$- v^*_2$};
\draw [->] (0,0) -- (0,2) node[above] {$-v^*_1$};
\draw [->] (0,0) -- (0,-2) node[below] {$v^*_1$};
\draw [->] (0,0) -- (2,-2) node[below] {$  -v_3^*$};
\draw [->] (0,0) -- (-2,2) node[above] {$v_3^*$};
\node at (1,1) {$\sigma^*_1$};
\node at (-1.5,0.5) {$\sigma^*_2$};
\node at (0.5,-1.5) {$\sigma^*_3$};
\end{tikzpicture}
 \\
\end{array}
\]
\caption{Toric fan of $\mathbb{P}^2$.  }
\end{figure}
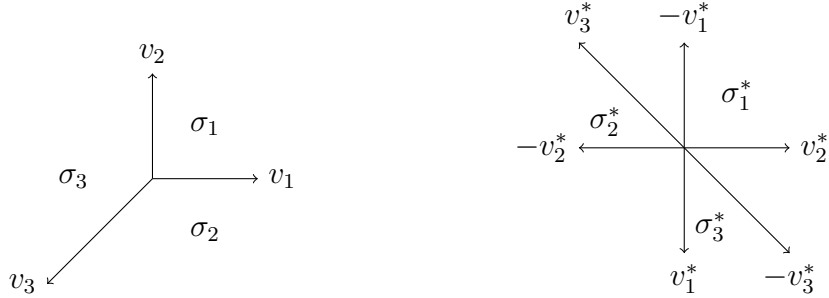

The fan  of $\mathbb{P}^2$ is specified by the vectors (see Appendix \ref{toricapp} for a brief introduction on toric geometry):
\be
 (v_1,v_2,v_3) = (  {\bf e}_1,{\bf e}_2, -{\bf e}_1- {\bf e}_2)    \qquad\qquad
 (v^*_1,v^*_2,v^*_3) =    (-{\bf e}^*_2 ,{\bf e}^*_1 ,{\bf e}^*_2- {\bf e}^*_1)
    \ee
The three vectors satisfy
\be
 v_1+v_2+v_3 =v^*_1+v^*_2+v^*_3 = 0  \label{relcp2}
\ee
 Comparing with the general toric formula
\be
v_{\ell-1} +v_{\ell+1}-h_\ell\, v_\ell=0
\ee
we conclude that $h_\ell=-C_{\ell \ell}=-1$. The non-trivial intersection numbers are
\be
D_\ell\cdot D_\ell=  D_\ell \cdot D_{\ell+1}=1
\ee
In addition, the relation  (\ref{relcp2}) determines the weights of the homogeneous coordinates in the description of the toric manifold
 as the quotient of $\mathbb{C}^{3} \setminus \{0\}$ by the equivalence relation
\be
 (y_1,y_2,y_3)\sim    (\lambda\, y_1,\lambda \, y_2, \lambda \, y_3)
\ee
In table \ref{cp2obs} we display the local coordinates $(z_1^\ell,z_2^\ell)$ in each chart $\ell$ and the corresponding equivariant parameters $(\epsilon_1^\ell,\epsilon_2^\ell)$.
\par For the zero form part of $w^{\ell}$ at the origin $z_1^{\ell'}=z_2^{\ell'}=0$
of chart $\ell'$ one finds
\be
\left[ w^{\ell} \right]_{\ell'} =
 \left(
\begin{array}{ccc}
\epsilon_1  & 0   & \epsilon_1 -\epsilon_2   \\
\epsilon_2  & \epsilon_2-\epsilon_1 & 0  \\
0  & -\epsilon_1  & -\epsilon_2
\end{array}
\right)
\ee
 We collect the geometric data in Table \ref{cp2obs}.
 Finally the K\"ahler form on $\mathbb{P}^2$  is $w=\alpha (w^1+w^2+w^3)$ with $\alpha$ a real positive number giving the volume of the manifold which was normalized to 1 in section 2.2.

  \begin{table}[!h]
  \bea
\left.
\begin{array}{|c|c|c|c|c|}
\hline
    \ell &    (z_1^\ell,z_2^\ell) &   (\epsilon^\ell_1,\epsilon^\ell_2)  &     a^{ \ell }   &   \Omega_\ell \\
  \hline
  1 &    \left( {y_1\over y_3},{y_2 \over y_3} \right) &   (\epsilon_1,\epsilon_2) &
   \mathfrak{a} +k^1 \epsilon_1 +k^2 \epsilon_2 &
  x^1 \epsilon_1+ x^2 \epsilon_2 +x^{12}\, \epsilon_1\, \epsilon_2 +T  \\
  2&   \left( {y_2\over y_1} , {y_3 \over y_1} \right)  &    (\epsilon_2-\epsilon_1,-\epsilon_1 )  &
    \mathfrak{a} +k^2 (\epsilon_2-\epsilon_1) -k^3 \epsilon_1  &
  x^2 (\epsilon_2-\epsilon_1)- x^3 \epsilon_1+x^{23}\, (\epsilon_1^2-\epsilon_1\, \epsilon_2)+T   \\
  3 &    \left( {y_3\over y_2} , {y_1 \over y_2} \right)   & (-\epsilon_2, \epsilon_1-\epsilon_2)    &
  \mathfrak{a} -k^3 \epsilon_2+k^1 (\epsilon_1-\epsilon_2)     &  x^1(\epsilon_1-\epsilon_2)
  -x^3 \epsilon_2+x^{13}\, (\epsilon_2^2-\epsilon_1\, \epsilon_2) +T \\
   \hline
\end{array}
\right.   \nonumber
\eea
\caption{ Observables on $\mathbb{P}^2$.  }
\label{cp2obs}
\end{table}

\subsection{Donaldson invariants}

\par To compare with the existing literature we set the Donaldson variables to
\be
x^3=z \quad, \quad  x^{13}=x^{23}=-x^{12}=x    \quad, \quad T=x \epsilon_1 \epsilon_2  \quad, \quad
x^1=x^2=0
\ee
or equivalently
\be
\Omega_\ell = \left( 0, -z \epsilon_1 +x (\epsilon_1)^2,  -z \epsilon_2 +x (\epsilon_2)^2 \right)
\ee
The classical contribution to the partition function becomes
\begin{equation}
   Z_{\rm class}  =   q^{\Delta_k+\frac{c_1^2}{4}}e^{\Delta^{\Omega}}
\end{equation}
where
\bea
\Delta_k &=& -\frac{r^2}{4} \qquad , \qquad  r=k^1+k^2+k^3   \qquad , \qquad c_1^2=r ~{\rm mod}~ 2
\eea
As we discussed before, for $\mathbb{P}^2$ there are only poles of order one. Let us consider the contribution of
 an orbit defined by the triplets $\{ k^1,k^2,k^3 \}$ of positive integers\footnote{Triplets involving a vanishing $k^\ell$ lead to a regular $Z_{\rm full}$ with vanishing residue.}.
 Such orbit contains 4 terms with $\kappa \geq 0$.
 The types of orbits  can be grouped as follows:
\begin{itemize}
\item{ Unstable orbits: they are  generated from a triplet  $(k_1,k_2,k_3)$ violating one of the triangle inequalities, let us say $k_1+k_2<k_3$.
  There are four contributions  $(\pm k_1,\pm k_2,k_3)$ which cancel against each other in pairs. }

  \item{ Stable orbits: they are  generated from a triplet,
satisfying the triangle inequalities $k_i+k_j>k_k$, and its flips. They
 contribute  with a factor $2-6=-4$.
 }

  \item{ Semistable orbits: they are  generated from a triplet
 $(k_1,k_2,k_3)$  saturating one of the triangle inequalities, let us say $k_1+k_2=k_3$.
   A contribution is missing, so that their contribution is weighted by a factor $2-4=-2$.
 }

\end{itemize}

\begin{figure}
	\includegraphics[width=1 \textwidth]{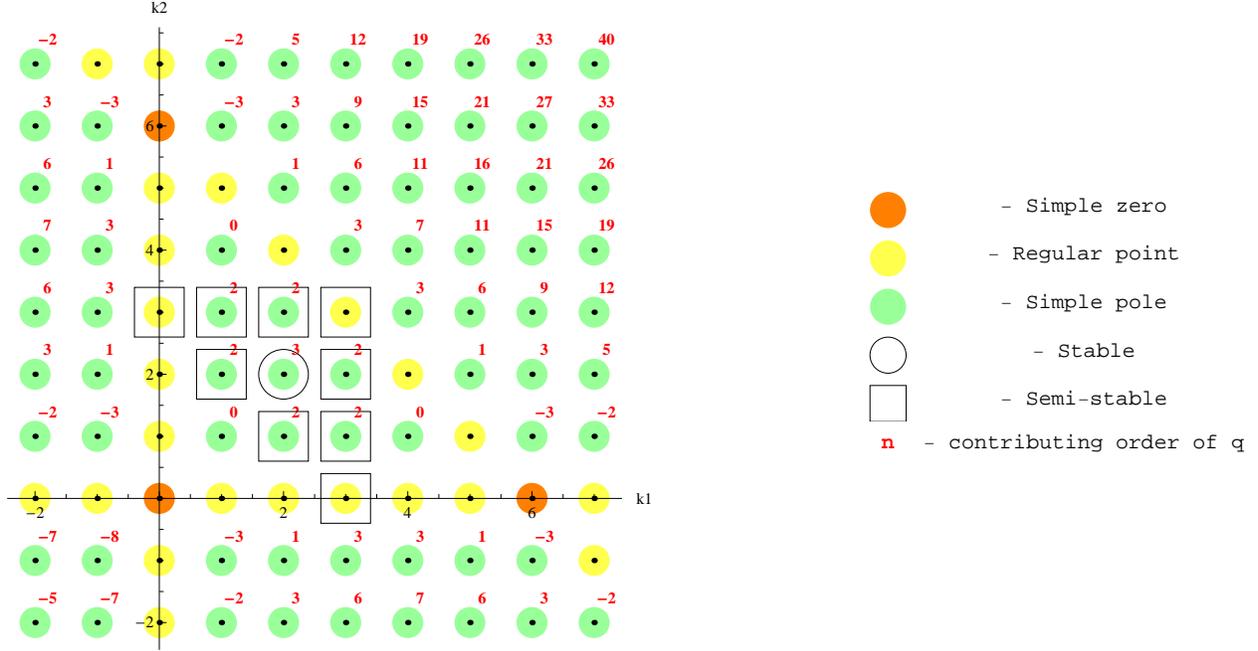}
	\caption{Poles of $Z_{\rm full}$ (with $r=k^1+k^2+k^3=6$) in the U(2) theory on $\mathbb{P}^2$.} \label{CP2zerosandpoles}
\end{figure}

If one of the $k_i$'s is zero, the number of flips is two times less, but it is compensated with the additional factor $1/2^\zeta$ from (\ref{zeta}).
\par In Fig \ref{CP2zerosandpoles}, we display the poles in the $a$-plane for $r=6$. Stable points are points inside the triangle, while semi-stable ones lie at the boundary of the triangle.  The partition function can then be written as
\be
   Z_M(\epsilon_a,q)=-4\sum_{{\bf k} \geq 0} Z_{\rm stable\, point}( {\bf k},\epsilon_a,q) - 2 \sum_{{\bf k} \geq 0} Z_{\rm semi \,stable\,point}( {\bf k},\epsilon_a,q)
     \ee
The contributions of each orbit can be computed by using the abstruse
duality \eqref{signflip}. Indeed, since $Z_{\rm full}$ has at most a single pole, its residue can be written as
\bea
 {\rm Res}_{ \mathfrak{a}=0} Z_{\rm full}(\mathfrak{a} ,{\bf k},\epsilon_a,q) &=&  {\rm Res}_{\hat{\mathfrak{a}}=0} Z_{\rm full}( \hat{\mathfrak{a}},{\bf k},\epsilon_a,q) \nn\\
 &=&   q^{\hat \Delta_k+\frac{c_1^2}{4}} e^{\hat \Delta_\Omega}  \widehat{Z}_{\rm 1loop}({\bf k},\epsilon_1,\epsilon_2)   \prod_\ell  Z_{\rm inst, \mathbb{C}^2 } (\hat{a}^\ell, \epsilon_1^\ell , \epsilon_2^\ell, q^\ell) \label{zzhat}
\eea
with
 \bea
 \hat \Delta_k &=&  \ft14 \left(2 k^1 k^2+2k^2 k^3+2 k^3 k^1-(k^1)^2-(k^2)^2-(k^3)^2\right)\nn\\
\widehat{Z}_{\rm 1loop}({\bf k},\epsilon_1,\epsilon_2)  &=& \prod_{m,n,p}    ( m \epsilon_1+n \epsilon_2)^{-d_{mn}^{\bf k} }
\label{zonetty1}
\eea
  with  $d^{\bf k}_{mnp }$ the expansion coefficients of the one-loop dual character
   \be
\widehat{\chi}_{\rm M}({\bf k} | t_1,t_2 )=   \sum_{\ell=1}^\chi {     \left(t_1^\ell \right)^{k^{\ell} } \, \left(t_2^\ell \right)^{-k^{\ell+1}}
+   \left(t_1^\ell \right)^{-k^{\ell} } \, \left(t_2^\ell \right)^{k^{\ell+1}}
\over (1- t_1^\ell)(1-t_2^\ell)   }-1= \sum_{m,n} d^{\bf k}_{m n } \, t_1^m t_2^n
\label{charonetty2}
\ee
 where the $-1$ removes the zero eigenvalue associated to the residue. It is interesting to observe that the stable orbits are characterised
 by polynomials with  expansion coefficients $d^{\bf k}_{m n }$ all positive, while semi-stable ones correspond to characters with
 all positive coefficients except one. The same pattern is observed for higher rank theories.
 We stress the fact that, although the two sides of equation (\ref{zzhat}) lead to the same results, the right hand side of (\ref{zzhat}) is easier  to evaluate  since $\hat{\Delta}_k$ is always positive and the instanton part is regular at $\hat{a}^{(mn)}$.

In order to compute the Donaldson invariants up to order $q^2$, for instance, it is enough to take the sum over $\textbf{k}$ with $k^\ell \leq 3$. The non-trivial contributions come from the orbits
\bea
c_1=1\qquad && {\bf k}=\left\{  (1,1,1), (1,2,2),\ldots.  \right\}+{\rm permutations} \nn\\
c_1=0\qquad && {\bf k}=\left\{  (1,1,2),  (1,2,3)  ,\ldots.  \right\} +{\rm permutations}
\eea
where the terms ``+permutations'' in the above formula are referred to the permutations of ${\bf k}$.
Setting
\be
Z_{c_1^2}(\mathbb{P}^2,q)=-{1\over 4} Z_M \big|_{c_1^2=k^1+k^2+k^3 ~{\rm mod}~2}
\ee
one finds
  \begin{eqnarray}
         Z_{c_1^2=1}(\mathbb{P}^2,q,\epsilon_1,\epsilon_2) &=& q \, e^{\frac{1}{4}(x \epsilon_1 \epsilon_2 - z(\epsilon_1+\epsilon_2))} +  q^2 \big(\frac{e^{\frac{1}{4}(x \epsilon_1 (\epsilon_2-\epsilon_1)-z(3\epsilon_1+\epsilon_2))}}{\epsilon_1^2 (\epsilon_1-\epsilon_2)\epsilon_2} +  \\
         &+& \frac{(\epsilon_1^2+\epsilon_2^2)e^{\frac{1}{4}(x \epsilon_1 \epsilon_2 -z(\epsilon_1+\epsilon_2))}}{2\epsilon_1^2 (\epsilon_1-\epsilon_2)^2 \epsilon_2^2} +
         \frac{e^{\frac{1}{4}(x \epsilon_1(4\epsilon_1+\epsilon_2)-z(5 \epsilon_1+\epsilon_2))}}{2\epsilon_1^2 (\epsilon_1-\epsilon_2)^2} + \nonumber \\
         &+& \frac{e^{\frac{1}{4}(x\epsilon_2(\epsilon_1+4\epsilon_2)-z(\epsilon_1+5\epsilon_2))}}{2(\epsilon_1-\epsilon_2)^2 \epsilon_2^2} + \frac{e^{\frac{1}{4}(x(\epsilon_1-\epsilon_2)\epsilon_2-z(\epsilon_1+3\epsilon_2))}}{\epsilon_1 (\epsilon_2-\epsilon_1)\epsilon_2^2} -  \nonumber \\
         &-& \frac{e^{\frac{1}{4}(x \epsilon_1 \epsilon_2-z(\epsilon_1+\epsilon_2))}}{\epsilon_1(\epsilon_1-\epsilon_2)^2\epsilon_2} - \frac{e^{-\frac{3}{4}z(\epsilon_1+\epsilon_2)-\frac{1}{4}x(\epsilon_1^2-7\epsilon_1\epsilon_2+\epsilon_2^2)}}{\epsilon_1(\epsilon_1-\epsilon_2)^2\epsilon_2} \big) + \ldots \nonumber
\end{eqnarray}
\begin{eqnarray}
Z_{c_1^2=0}(\mathbb{P}^2,q,\epsilon_1,\epsilon_2)&=&q \big(\frac{e^{-\frac{1}{4}x\epsilon_1(\epsilon_1-4\epsilon_2)-z \epsilon_2}(\epsilon_1-2\epsilon_2)}{2(\epsilon_1-\epsilon_2)\epsilon_2} + \frac{e^{-z\epsilon_1+x \epsilon_1 \epsilon_2-\frac{x \epsilon_2^2}{4}}(2\epsilon_1-\epsilon_2)}{2\epsilon_1(\epsilon_1-\epsilon_2)} -\nonumber \\
&-& \frac{e^{-\frac{1}{4}x(\epsilon_1+\epsilon_2)^2}(\epsilon_1+\epsilon_2)}{2\epsilon_1 \epsilon_2}\big) +  \ldots \nonumber \\
\end{eqnarray}

  In the limit of  $\epsilon_1, \, \epsilon_2 \rightarrow 0$ one recovers the standard non-equivariant Donaldson invariants
 \bea
  \label{DonCP2}
 Z_{c_1^2=1}(\mathbb{P}^2,q)&=& q+{q^2 \over 16} \left( 19  \frac{x^2}{2!}+ 5 \frac{x z^2}{2!}+3\frac{z^4}{4!} \right) +\ldots\\
  Z_{c_1^2=0}(\mathbb{P}^2, q)&=& -{3q z\over 2}+q^2  \left( -  \frac{13}{ 8}\,\frac{x^2 z}{2!} -{x z^3 \over 3!}+\frac{z^5}{5!} \right)+\ldots
 \eea

\section{Gauge theories on $\mathbb{F}_n$}
\label{section6}
\setcounter{equation}{0}
In this section we consider $SU(2)$ gauge theories on $F_n$.

\subsection{Geometric data}

 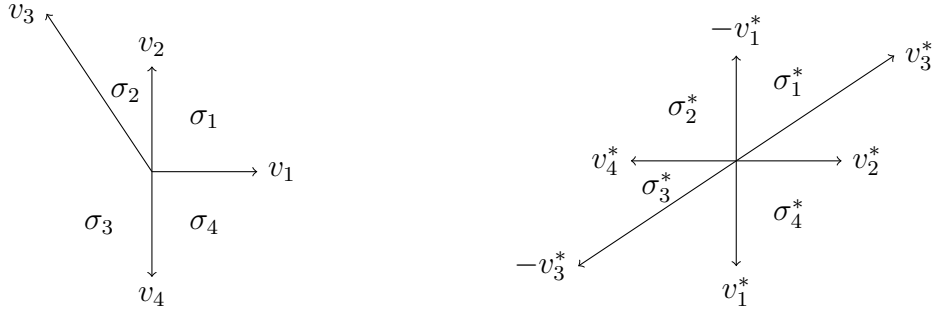
\begin{figure}[!h]
\centering
\[
 \begin{array}{ccc}
\begin{tikzpicture}[scale=0.7]
\draw [->] (0,0) -- (2,0) node[right] {$v_1$};
\draw [->] (0,0) -- (0,2) node[above] {$v_2$};
\draw [->] (0,0) -- (-2,3) node[left] {$v_3$};
\draw [->] (0,0) -- (0,-2) node[below] {$v_4$};
\node at (1,1) {$\sigma_1$};
\node at (-0.5,1.5) {$\sigma_2$};
\node at (-1,-1) {$\sigma_3$};
\node at (1,-1) {$\sigma_4$};
\end{tikzpicture}
& ~~~~~~~~~~~~~~&
\begin{tikzpicture}[scale=0.7]
\draw [->] (0,0) -- (2,0) node[right] {$v_2^* $};
\draw [->] (0,0) -- (0,-2) node[below] {$v_1^* $};
\draw [->] (0,0) -- (0,2) node[above] {$-v_1^* $};
\draw [->] (0,0) -- (-2,0) node[left] {$v_4^* $};
\draw [->] (0,0) -- (3,2) node[right] {$v_3^* $};
\draw [->] (0,0) -- (-3,-2) node[left] {$-v_3^* $};
\node at (1,1.5) {$\sigma_1^*$};
\node at (-1,1) {$\sigma_2^*$};
\node at (-1.5,-0.5) {$\sigma_3^*$};
\node at (1,-1) {$\sigma_4^*$};
\end{tikzpicture} \\
\end{array}
\]
\caption{The toric fan of $\mathbb{F}_n$.
$\sigma_\ell$ labels the cone of dimension two relative to the $\ell$-th $\mathbb{C}^2$ coordinates patch.}
\end{figure}

The four vectors of the toric fan of  $\mathbb{F}_n$  satisfy the relations
\be
  v_1+v_3 +n v_4 =  v_2+v_4=0    \label{reffn}
\ee
leading to the identification
 \be
 (y_1,y_2,y_3,y_4) \sim    (\lambda \, y_1, \lambda' \, y_2, \lambda \, y_3, \lambda' \lambda^n \, y_4)
 \ee
 The non-trivial intersection numbers following from (\ref{reffn}) are
 \be
 C_{44}=-C_{22}=n \, , \qquad C_{\ell,\ell \pm 1}=1
 \ee
The zero form part of $w^{\ell'}$ at the origin of chart $\ell$ can be found as
\be
\left[ w^{\ell'} \right]_\ell =
 \left(
\begin{matrix}
\epsilon_1  & 0    & 0 & \epsilon_1   \\
\epsilon_2  & \epsilon_2+n \epsilon_1  & 0 & 0 \\
0  & -\epsilon_1  & -\epsilon_1 & 0  \\ 0 & 0& -\epsilon_2-n \epsilon_1 & -\epsilon_2
\end{matrix}
\right)
\ee
The coefficients of the non-equivariant curvature are
 \bea
 r &=&  k^{1}+k^{3}  -n\,  k^2 \nn\\
 r' &=&    k^{2}+k^{4}
  \eea
 We collect the geometric  data in Table \ref{ttoric4}.

     \begin{table}[!h]
  \bea
\left.
\begin{array}{|c|c|c|c|c|c|}
\hline
  \ell &  (z_1^\ell,z_2^\ell) &    (\epsilon^\ell_1,\epsilon^\ell_2) & a^\ell & \Omega^\ell  \\
  \hline
  1 &  \left( {y_1 \over y_3} , {y_2 y_3^n \over y_4} \right) &   (\epsilon_1,\epsilon_2) & \mathfrak{a} +k^1 \epsilon_1 +k^2 \epsilon_2   & x^1 \epsilon_1 + x^2 \epsilon_2 + x^{12} \epsilon_1 \epsilon_2  \\
  2&  \left( {y_1^n y_2 \over y_4} , {y_3 \over y_1 } \right) &  (n\epsilon_1 + \epsilon_2,-\epsilon_1 ) &\mathfrak{a} +k^2(n \epsilon_1+\epsilon_2) -k^3 \epsilon_1  & x^2(\epsilon_2+n \epsilon_1) - \epsilon_1 x^3 - x^{23} (\epsilon_1 \epsilon_2+n \epsilon_1^2)   \\
  3&  \left( {y_3\over y_1} , {y_4 \over y_1^n y_2} \right) &   (-\epsilon_1,-n \epsilon_1 -\epsilon_2)   & \mathfrak{a} -k^3 \epsilon_1 -k^4 (n\epsilon_1+\epsilon_2)  & -x^3 \epsilon_1 -x^4 (\epsilon_2+n \epsilon_1) + x^{34}(\epsilon_1 \epsilon_2+n\epsilon_1^2)\\
  4 &  \left(  {y_4 \over y_2  y_3^n} , {y_1 \over y_3}  \right)  &  (-\epsilon_2,\epsilon_1 ) & \mathfrak{a} -k^4 \epsilon_2 +k^1 \epsilon_1 & x^1 \epsilon_1 - x^4 \epsilon_2 - x^{14} \epsilon_1 \epsilon_2 \\
  \hline
\end{array}
\right.   \nonumber
\eea
\caption{ Geometric data for $  \mathbb{F}_n$.    }
\label{ttoric4}
\end{table}

 \subsection{The Donaldson invariants}
\begin{figure}
\includegraphics[width=0.5 \textwidth]{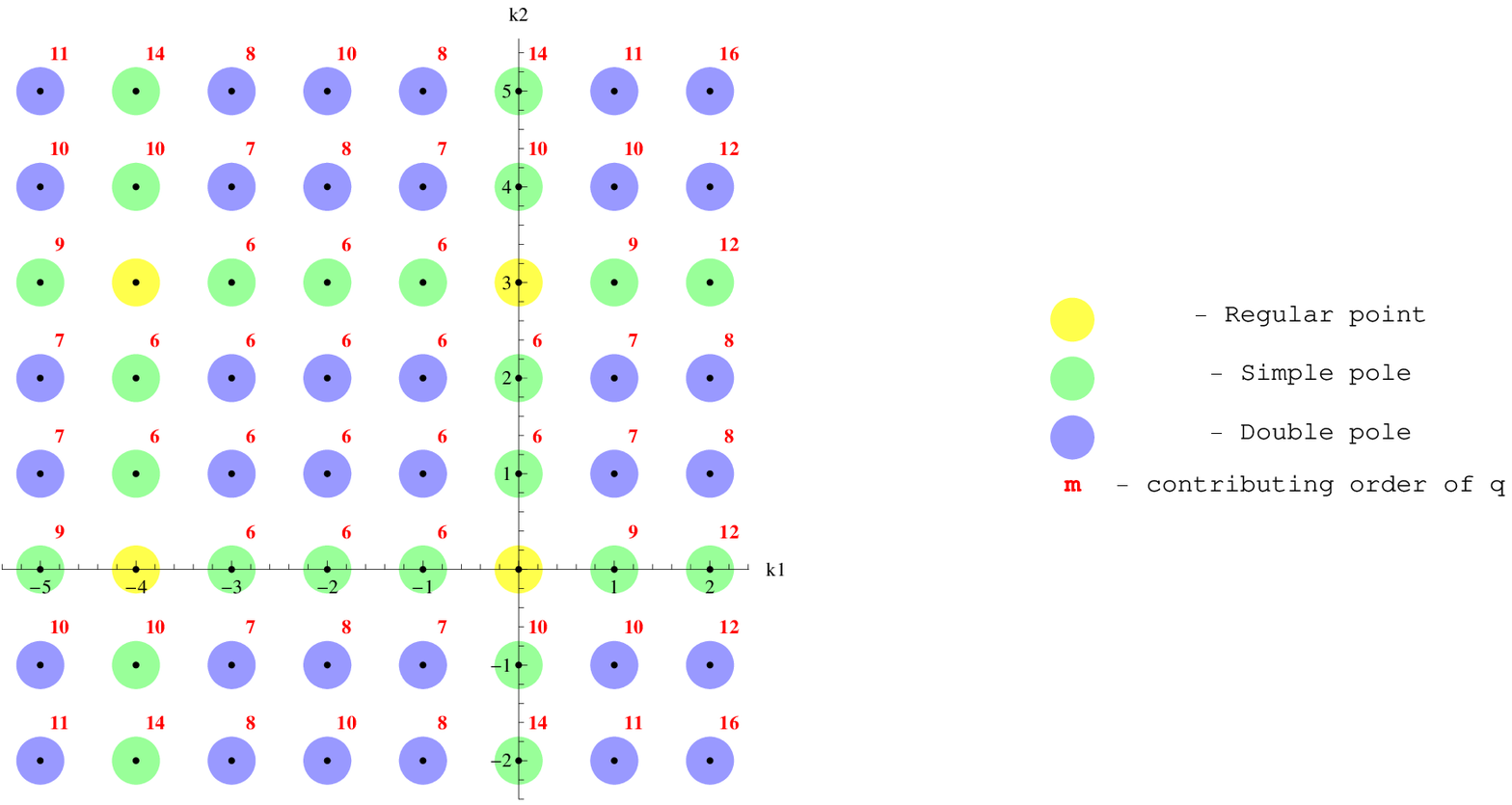} \quad
\includegraphics[width=0.5 \textwidth]{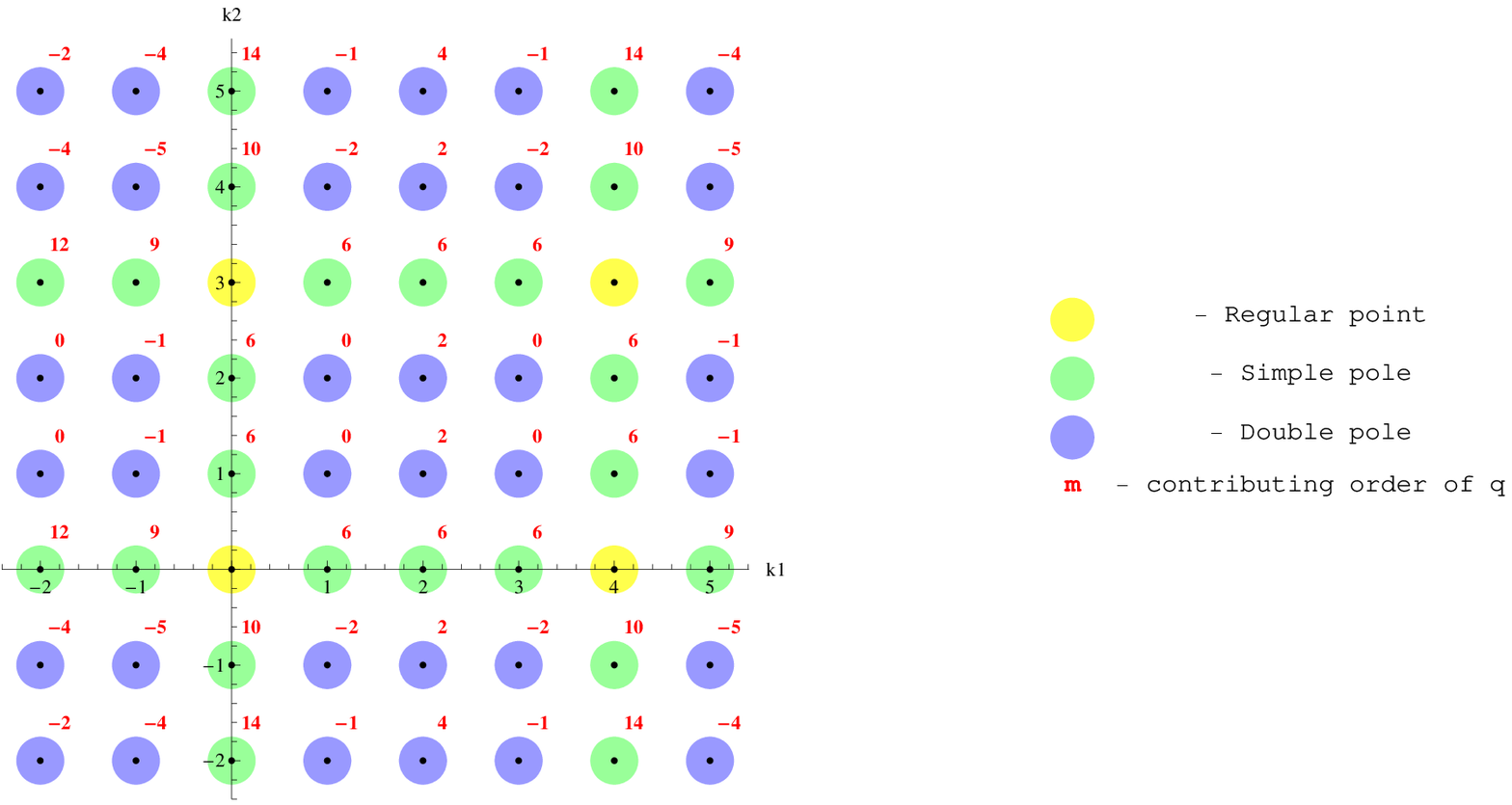}
\caption{Poles of $Z_{\rm full}$ in the U(2) theory on $\mathbb{F}_0$ for $(r,r')$:  a) (-4,3), b) (4,3) } \label{Fnpolesneg}
\end{figure}

We take the K\"ahler form to be $\omega=\alpha w_1+\beta w_2$ with $\alpha-n \beta >0$,
and set the parameters of the observables as
\begin{eqnarray}
x^1=x^2=0, \, x^3=z_1, \, x^4=z_2 \nn \\
x^{12}=x^{23}=x^{14}=0, x^{34}=x   \nn
\end{eqnarray}
Or equivalently
\be
\Omega_\ell = \left( 0, -z_1 \epsilon_1 , - z_1 \epsilon_1  -z_2( \epsilon_2+n \epsilon_1) +x
(\epsilon_1\epsilon_2 + n \epsilon_1^2)  ,  -z_2 \epsilon_2 \right)
\ee
The maximal order of a pole is $\chi-2=2$ and any $k^\ell$ becoming zero decreases its order by one, see Fig \ref{Fnpolesneg} for an example.

 The classical contribution to the partition function is
\begin{equation}
   Z_{\rm class}  =   q^{\Delta_k+\frac{c_1^2}{4}}e^{\Delta^{\Omega}}
\end{equation}
where we choose the first Chern class to be $c_1=(r \, {\rm mod \, }2 )w_3+(r' \, {\rm mod \,}2) w_4$, so that
\bea
\Delta_k &=& {1\over 4} (2\, r\, r'+n \,r'^2) \nn\\
    c_1^2 &=& 2 (r \, {\rm mod \, }2 ) (r' \, {\rm mod \, }2 )+n(r' \, {\rm mod \, }2 )^2
\eea

 \subsection*{ Hirzebruch surface $\mathbb{F}_0$}

In order to compute the Donaldson invariants up to $q^2$ it is enough to take values up to $k_i=2$. To be more precise, the non-trivial contributions come from the orbits  \bea
c_1=(0,0) \quad && {\bf k}=\left\{  (0,1,2,1), (1,0,1,2),(1,2,1,0), (2,1,0,1),(1,1,1,1),
(1,2,1,2),(2,1,2,1) \right\}  \nn\\
c_1=(0,1) \quad && {\bf k}=\left\{  (1,0,1,1), (1,1,1,0),(1,1,1,2), (1,2,1,1),(2,0,2,1),
(2,1,2,0) \right\}  \nn\\
c_1=(1,1) \quad && {\bf k}=\left\{  (0,1,1,2), (0,2,1,1), (1,0,2,1),(1,1,0,2), (1,1,2,0),(1,2,0,1) \right. \nn\\
 && \left. \qquad ,(2,0,1,1), (2,1,1,0) \right\}
\eea

\par For the equivariant Donaldson invariants one gets
\begin{eqnarray}
   Z_{c_1=(0,0)}(\mathbb{F}_0,q,\epsilon_1,\epsilon_2)&=&q \big(\frac{(\epsilon_1-\epsilon_2) e^{-\frac{1}{4}x(\epsilon_1-\epsilon_2)^2-z_2 \epsilon_2}}{2\epsilon_1 \epsilon_2} - \frac{(\epsilon_1-\epsilon_2) e^{-\frac{1}{4}x(\epsilon_1-\epsilon_2)^2-z_1 \epsilon_1}}{2\epsilon_1 \epsilon_2}  + \\
   &+&\frac{(\epsilon_1+\epsilon_2) e^{-\frac{1}{4}x(\epsilon_1-\epsilon_2)^2-z_1 \epsilon_1-z_2 \epsilon_2}}{2\epsilon_1 \epsilon_2} -\frac{(\epsilon_1+\epsilon_2) e^{-\frac{1}{4}x(\epsilon_1-\epsilon_2)^2}}{2\epsilon_1 \epsilon_2} \big) + \ldots \nonumber
\end{eqnarray}
\begin{equation}
    Z_{c_1=(1,0)}(\mathbb{F}_0,q,\epsilon_1,\epsilon_2)=q \big(-\frac{e^{-\frac{1}{4}(x(\epsilon_1-\epsilon_2)^2+4 z_2 \epsilon_2+z_1(2\epsilon_1-\epsilon_2))}}{\epsilon_2} +\frac{e^{-\frac{x \epsilon_1^2}{4}-\frac{1}{4}z_1(2\epsilon_1+\epsilon_2)}}{\epsilon_2} \big)+\ldots
\end{equation}
\begin{eqnarray}
    Z_{c_1=(1,1)}(\mathbb{F}_0,q,\epsilon_1,\epsilon_2)&=& q^{2} \big( \frac{e^{-\frac{1}{4}(x \epsilon_1^2+2z_1\epsilon_1+3z_2\epsilon_1+z_1\epsilon_2)}}{2 \epsilon_1 \epsilon_2 (\epsilon_1-\epsilon_2)} + \frac{e^{\frac{1}{4}(-x(\epsilon_1-2\epsilon_2)^2+3z_1(-2\epsilon_1+\epsilon_2)+z_2(\epsilon_1-4\epsilon_2))}}{2 \epsilon_1 \epsilon_2 (\epsilon_1-\epsilon_2)} - \nonumber \\
    &-& \frac{e^{\frac{1}{4}(-x(\epsilon_2-2\epsilon_1)^2+z_1(\epsilon_2-4\epsilon_1)+3z_2(\epsilon_1-2\epsilon_2))}}{2 \epsilon_1 \epsilon_2 (\epsilon_1-\epsilon_2)} - \frac{e^{-\frac{1}{4}(x \epsilon_2^2 +3 z_1 \epsilon_2 +z_2 (\epsilon_1 +2\epsilon_2))}}{2 \epsilon_1 \epsilon_2 (\epsilon_1-\epsilon_2)} - \nonumber \\
    &-& \frac{e^{\frac{1}{4}(-x(\epsilon_1-\epsilon_2)^2-3z_1(2\epsilon_1+\epsilon_2)+z_2(\epsilon_1-2 \epsilon_2))}}{2 \epsilon_1 \epsilon_2 (\epsilon_1+\epsilon_2)}+\frac{e^{\frac{1}{4}(-4x \epsilon_1^2-z_1(4\epsilon_1+\epsilon_2)+3z_2 \epsilon_1)}}{2 \epsilon_1 \epsilon_2 (\epsilon_1+\epsilon_2)} - \\
    &-& \frac{e^{\frac{1}{4}(-x(\epsilon_1-\epsilon_2)^2+z_1(-2\epsilon_1+\epsilon_2)-3z_2(\epsilon_1+2\epsilon_2))}}{2 \epsilon_1 \epsilon_2 (\epsilon_1+\epsilon_2)} + \frac{e^{\frac{1}{4}(-4x\epsilon_2^2+3z_1\epsilon_2-z_2(\epsilon_1+4\epsilon_2))}}{2 \epsilon_1 \epsilon_2 (\epsilon_1+\epsilon_2)} \big)+\ldots \nonumber
\end{eqnarray}
  In the non equivariant limit $\epsilon_1, \, \epsilon_2 \to 0$ one finds
 \bea
  \label{DonF0}
 Z_{c_1=(0,0)}(\mathbb{F}_0,q)&=& q \left(-z_1-z_2\right)-\frac{1}{960} q^2 \left(z_1+z_2\right) \left(525 x^2+20 x \left(z_1^2+8 z_2 z_1+z_2^2\right)  \right.\nn\\
 &&\left.  -4 \left(z_1^4-6 z_2 z_1^3+16
   z_2^2 z_1^2-6 z_2^3 z_1+z_2^4\right)\right) +\ldots\nn\\
  Z_{c_1=(1,0)}(\mathbb{F}_0,q)&=& \frac{q^2 \left(2160 x^2 z_2-20 z_1^3 \left(7 x-8 z_2^2\right)-240 x z_2 z_1^2-60 x z_1 \left(17 x-16 z_2^2\right) +31 z_1^5-140 z_2
   z_1^4\right)}{3840}\nn\\
&&   +q \left(z_2-\frac{z_1}{2}\right) +\ldots\nn\\
 Z_{c_1=(1,1)}(\mathbb{F}_0,q)&=& \frac{1}{48} q^{2} \left(z_1+z_2\right) \left(-6 x+13 z_1^2-22 z_2 z_1+13 z_2^2\right)+\ldots
  \eea
  The results for $c_1=(0,1)$ and $c_1=(1,1)$ perfectly match those obtained using the wall crossing formulae. Indeed, in these two cases,
  an empty room exists  and the contribution of every orbit is equal to a contribution of its flip with $r r' \leq 0$ with an additional factor $(-4)$ for the stable points and $(-2)$ for the semistable ones in agrement with the wall crossing results. On the other hand, in the case $c_1=(0,0)$ there is no empty room,  and the contribution of the orbits $\textbf{k}=(1,1,1,1)$ is not proportional to a contribution of any of its flips satisfying the condition $r r'\leq 0$.

  \subsection*{ Hirzebruch surface $\mathbb{F}_1$}

Again, in order to compute the Donaldson invariants up to $q^2$ it is enough to take values of the gauge fluxes up to $k_i=2$. The non-trivial contributions come from the orbits
 \bea
c_1=(0,0) \quad && {\bf k}=\left\{  (1,0,1,2),(1,1,2,1), (2,1,1,1)  \right\}  \nn\\
c_1=(0,1) \quad &&  {\bf k}=\left\{  (1,0,1,1), (1,1,2,2),(1,2,1,1), (2,0,2,1),(2,1,1,2)  \right\}  \nn\\
c_1=(1,0) \quad &&  {\bf k}=\left\{  (1,1,1,1)  \right\}  \nn\\
c_1=(1,1) \quad &&  {\bf k}=\left\{  (1,0,2,1), (1,1,1,2),(1,2,2,1), (2,0,1,1),(2,2,1,1)
 \right\}  \nn
\eea
For the equivariant Donaldson invariants one gets
\begin{eqnarray}
    Z_{c_1=(0,0)}(\mathbb{F}_1,q,\epsilon_1,\epsilon_2)&=&\frac{q}{2} (\frac{e^{-\frac{1}{4}x (\epsilon_1-\epsilon_2)^2-z_2\epsilon_2}(\epsilon_1-\epsilon_2)}{\epsilon_1 \epsilon_2}+\frac{e^{-z_1\epsilon_1-z_2(\epsilon_1+\epsilon_2)-\frac{x \epsilon_2^2}{4}}(2\epsilon_1+\epsilon_2)}{\epsilon_1(\epsilon_1+\epsilon_2)} -\nonumber \\
    &-& \frac{e^{z_1 \epsilon_2-\frac{1}{4}x(\epsilon_1+2\epsilon_2)^2}(\epsilon_1+2\epsilon_2)}{\epsilon_2(\epsilon_1+\epsilon_2)}\big)+\ldots
\end{eqnarray}
\begin{equation}
    Z_{c_1=(0,1)}(\mathbb{F}_1,q,\epsilon_1,\epsilon_2)=q \, e^{\frac{1}{4}(-(z_1+z_2)\epsilon_1+(z_1-2z_2)\epsilon_2-x\epsilon_2^2)}+\ldots
\end{equation}
\begin{equation}
    Z_{c_1=(1,0)}(\mathbb{F}_1,q,\epsilon_1,\epsilon_2)=q\frac{e^{z_1 \epsilon_1}-1}{\epsilon_1}e^{-z_1\epsilon_1-\frac{x \epsilon_2^2}{4}-\frac{1}{4}z_2(\epsilon_1+2\epsilon_2)}+\ldots
\end{equation}
\begin{eqnarray}
    Z_{c_1=(1,1)}(\mathbb{F}_1,q,\epsilon_1,\epsilon_2)&=&q^2 \big( -\frac{e^{-\frac{x\epsilon_2^2}{4}-z_2(\epsilon_1+\epsilon_2)-\frac{3}{4}z_1(3\epsilon_1+\epsilon_2)}}{2\epsilon_1(\epsilon_1+\epsilon_2)} + \frac{e^{\frac{1}{4}(-x\epsilon_2^2+z_1(\epsilon_2-\epsilon_1)-4z_2(\epsilon_1+\epsilon_2))}}{2\epsilon_1(\epsilon_1+\epsilon_2)} - \\
    &-& \frac{e^{\frac{1}{4}(-x(\epsilon_1-\epsilon_2)^2-4z_2\epsilon_2+z_1(\epsilon_2-3\epsilon_1))}}{2\epsilon_1\epsilon_2}+\frac{e^{\frac{1}{4}(z_1(\epsilon_1-3\epsilon_2)-x(\epsilon_1-\epsilon_2)^2-4z_2\epsilon_2)}}{2\epsilon_1\epsilon_2} - \nonumber \\
    &-& \frac{e^{\frac{1}{4}(-x(\epsilon_1+2\epsilon_2)^2)+z_1(\epsilon_1+3\epsilon_2)}(\epsilon_1+2\epsilon_2)}{2\epsilon_1\epsilon_2(\epsilon_1+\epsilon_2)}+ \frac{e^{-\frac{1}{4}x(\epsilon_1+2\epsilon_2)^2-\frac{3}{4}z_1(\epsilon_1-\epsilon_2)}(\epsilon_1+2\epsilon_2)}{2\epsilon_1\epsilon_2(\epsilon_1+\epsilon_2)} \big) +\ldots\nonumber
\end{eqnarray}

  In the non-equivariant limit $\epsilon_1, \epsilon_2\to 0$ one finds
 \bea
  \label{DonCP2}
 Z_{c_1=(0,0)}(\mathbb{F}_1,q)&=& -\ft32 q\left(z_1+z_2\right) \nn\\
  Z_{c_1=(0,1)}(\mathbb{F}_1,q)&=& q+ \frac{q^2}{384}  \left(228 x^2+60 x \left(z_1+z_2\right){}^2-29 z_1^4+12 z_2 z_1^3+18 z_2^2 z_1^2+12 z_2^3 z_1+3 z_2^4\right)\nn\\
    Z_{c_1=(1,0)}(\mathbb{F}_1,q)&=& q  z_1 \nn\\
 Z_{c_1=(1,1)}(\mathbb{F}_1,q)&=& -\frac{3}{2}q^2 z_1(z_1+z_2) 
 \eea

\section{SU(3) gauge theory on  $\mathbb{P}^2$}
\label{sectionsun}
 
In this section we compute the first instanton corrections to the Donaldson partition function for a theory with gauge group SU(3) living on $\mathbb{P}^2$. The sum over the gauge fluxes is spanned now by two triplets of integers $(k_1^\ell | k_2^\ell )$
 with $\ell=1,2,3$. Using the Weyl symmetry in each chart one can order the tuplet such that $k_1^\ell>k_2^\ell>0$. 
A scan over all tuplets with $k_\alpha^\ell \leq 2$ shows that the leading contributions arise at order $q^2$,  from the three tuplets
 \be
 (k_1^\ell | k_2^\ell ) =(1,2,2|1,1,1), (2,2,1|1,1,1), (2,1,2|1,1,1).  \label{ksu3s}
 \ee
  The three terms contribute at order $q^2$, with a $q^{\Delta_k+{1\over 3} }=q^{-6}$ coming from the classical part and a factor $q^{k_1^\ell k_1^\ell} =q^{2+2+4}$ coming from the instanton partition function in the three charts covering $\mathbb{P}^2$. An explicit evaluation of the residues leads to 
 \be\label{resea}
   \begin{aligned}
 {\rm Res}_{\mathfrak{a}_\alpha=0}\,   Z_{(1,2,2|1,1,1)} &=    -\frac{q^2 e^{\left(-\frac{z (\epsilon _1+ 2 \epsilon_2)}{3}+\frac{1}{3} x (\epsilon _1 \epsilon _2 - \epsilon _2^2)\right)}}{\epsilon _1 \left(\epsilon _1-\epsilon _2\right)}  +\ldots  \\
  {\rm Res}_{\mathfrak{a}_\alpha=0}\,   Z_{(2,1,2|1,1,1)} &= \frac{q^2 e^{ \left(-\frac{z (2\epsilon _1+\epsilon _2)}{3}+\frac{1}{3} x (\epsilon _1 \epsilon _2-\epsilon _1^2)\right)}}{\left(\epsilon _1-\epsilon _2\right) \epsilon _2} +\ldots  \\
   {\rm Res}_{\mathfrak{a}_\alpha=0}\,   Z_{(2,2,1|1,1,1)} &= -\frac{q^2 e^{\left(-\frac{2 z (\epsilon _1+\epsilon _2)}{3}+\frac{1}{3} x (5\epsilon _1 \epsilon _2-\epsilon _1^2-\epsilon _2^2)\right)}}{\epsilon _1 \epsilon _2}+\ldots
	   \end{aligned}
   \ee
 where the dots stands for higher instanton corrections.  
 
 The  results (\ref{resea}) can be alternatively found  by exploiting the abstruse duality. Indeed, since $Z_{\rm full}(\mathfrak{a}_\alpha,q)$ has a single pole in $\mathfrak{a}_1=\mathfrak{a}_2=0$, the residue receives
 contribution only from the leading term near the pole in each chart. Consequently, these contributions can be related using the abstruse duality (\ref{abduality2}) in each chart to the residue of the partition function  at $\hat{\mathfrak{a}}_\alpha=0$. More precisely, one finds
 \be
  {\rm Res}_{\mathfrak{a}_\alpha=0} Z_{\rm full}( \mathfrak{a}_\alpha, k_\alpha^\ell,\epsilon_a,q)= -{\rm Res}_{\hat{\mathfrak{a}}_\alpha=0} Z_{\rm full}(\hat{\mathfrak{a}}_\alpha , k_\alpha^\ell,\epsilon_a,q)  \label{zduality3}
\ee
  Notice that this relation holds for any tuplet such that the partition function exhibit a single pole at the origin. Remarkably, the right hand side of (\ref{zduality3}) can be easily evaluated to all orders in $q$. 
  The crucial simplification follows from the fact that  for $k^\ell_1>k_2^\ell>0$, the instanton partition function in the right hand side of (\ref{zduality3}) is regular at $\mathfrak{a}_\alpha=0$,  so one  has to deal with a residue of the much simpler one-loop part. The result can be written as
   \be
  {\rm Res}_{\hat{\mathfrak{a}}_\alpha=0} Z_{\rm full}(\hat{\mathfrak{a}}_\alpha , k_\alpha^\ell,\epsilon_a,q)  
  ={ Z_{\rm class}(\hat{a}^{\bf k}_\alpha , k_\alpha^\ell,\epsilon_a,q)   Z_{\rm inst}(\hat{a}^{\bf k}_\alpha , k_\alpha^\ell,\epsilon_a,q) \over \prod_{m,n} (m \epsilon_1+n \epsilon_2)^{d_{mn}^{\bf k} } } \label{resdu}
\ee
where $d^{\bf k}_{m n }$ follows from the dual one-loop character evaluated at $\mathfrak{a}_\alpha=0$ (the residue of the one-loop
partition function)
%
%
%
 \be
\widehat{\chi}_{\rm M}({\bf k} | t_1,t_2 )=   \sum_{\ell=1}^\chi  \sum_{u \neq v =1}^3 {  y_u^{{\bf k},\ell}  \, (y_v^{{\bf k},\ell})^{-1}
\over (1- t_1^\ell)(1-t_2^\ell)   }=2+ \sum_{m,n} d^{\bf k}_{m n } \, t_1^m t_2^n
\label{charonetty2}
\ee
 with the $2$ taking care of the residue and
\be
y_1^{{\bf k},\ell} =t_1^{k_1^\ell} \qquad \qquad
y_2^{{\bf k},\ell} =t_1^{k_2^\ell}\, t_2^{k_2^{\ell+1}} \qquad \qquad
y_3^{{\bf k},\ell}=t_2^{k_1^{\ell+1} }
\ee
One finds:
  \bea
\widehat\chi_M((1,2,2|1,1,1)|t_1,t_2) &=&  {1\over t_1}+{t_2\over t_1}   \nn\\
\widehat\chi_M((2,1,2|1,1,1)|t_1,t_2) &=&  {1\over t_2}+{t_1\over  t_2}   \nn\\
\widehat\chi_M((2,2,1|1,1,1)|t_1,t_2) &=& t_1+   t_2   \label{charsu3}
\eea
 Reading $d^{\bf k}_{mn}$ from (\ref{charsu3}) and plugging them into (\ref{resdu}) one reproduces  the denominators of (\ref{resea}) while the numerators come from the classical part of the partition function. The instanton contributes as 1 at this order, but an exact formula  follows from (\ref{resdu}) if all instanton terms are kept.  
 
 Putting together all pieces, we find that that each contribution in (\ref{resea}) is divergent in the limit $\epsilon_a\to 0$ but their sum is finite. Indeed, the sum of the three terms leads to
\be \label{ZM20}
Z_{M }(X,q) \sim  -\frac{ q^2}{18} (30 x + z^2)+ \ldots
\ee
 with dots denoting higher $\epsilon$'s and instanton contributions. Other contributions of tuplets in the Weyl orbits of these three terms will
 lead to the same results up to signs, so the total contribution of the three orbits will be again proportional to (\ref{ZM20}). 
 
  Finally we notice that the tuplets (\ref{ksu3s}) with $(\kappa_{1},\kappa_{2} )=(5,3)$ satisfy the slope stable conditions (\ref{slopecond})  and $(1,2,2|1,1,1)$, $(2, 1, 2|1,1,1)$, $(2,2,1|1,1,1)$ satisfy (\ref{Kool0}) as expected for a contributing term.

\vskip 0.2cm
\noindent {\large {\bf Acknowledgments}}
\vskip 0.02cm
G.B. and A.T. would like to thank P.~Putrov for interesting discussions and H.~Nakajima for encouragement. The research of G.B., F.F.\ and F.M.\ is partly supported by the INFN Iniziativa Specifica ST\&FI and by the PRIN project ``Non-perturbative Aspects Of Gauge Theories And Strings''.
The research of E.S.\ and A.T.\ is partly supported by the INFN Iniziativa Specifica GAST.
The work of A.T.\ is partially supported by the PRIN project ``Geometria delle varieta` algebriche''.

The work of M.R.\ has been partially supported by the funds awarded by the Friuli Venezia Giulia autonomous
Region Operational Program of the European Social Fund 2014/2020,
project ``HEaD - HIGHER EDUCATION AND DEVELOPMENT SISSA OPERAZIONE 3'', CUP G32F16000080009,
by INFN via Iniziativa Specifica GAST and by National Group of Mathematical Physics (GNFM-INdAM).
MR would like to thank Universit\'e de Gen\`eve for ospitality during the early stage of this project.
MR would like to thank the INFN Sezione di Roma ``TorVergata'' for ospitality during the early stage of this project. 

\begin{appendix}

 \section{Toric geometry} \label{toricapp}
\setcounter{equation}{0}
In this appendix we give a brief review of toric geometry.
  A  toric  variety $M$ of complex  dimension two is specified by a set of vectors $\{ v_\ell \} \in \mathbb{Z}^2$. Each cone $\sigma_\ell$ generated by $(v_\ell,v_{\ell+1})$ is isomorphic to a copy of $\mathbb{C}^2$, and the set of cones, the so called fan, defines a covering of $M$.
  The variety $M$ defined by the fan $\{ \sigma_\ell \}$ is compact if the fan covers the whole $\mathbb{R}^2$,  and the index $\ell$ is understood mod $\chi$, {\it i.e} $v_{\chi+1}=v_1$.
     The variety is smooth if any point in $\sigma_\ell\cup \mathbb{Z}^2$   can be written as a linear combination of $v_\ell$ and $v_{\ell+1}$ with positive integer coefficients.
We restrict ourselves to compact smooth varieties. The manifold can be equipped with $\chi$ global coordinates $(y_1,\ldots , y_\chi)$.
\par The vectors $v_\ell\in \mathbb{R}^2$ satisfy the relations
 \be
    v_{\ell-1}+v_{\ell+1} - h_{\ell} \, v_{\ell} = 0 \qquad \ell=1,\ldots  \chi  \label{constraint0}
\ee
We notice that only  $\chi-2$ of these relations are  independent.
To each ray $v_\ell$  we associate a divisor $D_\ell \sim \mathbb{P}^1$ defined as $y_\ell=0$
\par The integers $h_\ell$ specify the self-intersection numbers of the divisors in the toric geometry.
More precisely, the intersection pairing $ D_{\ell} \cdot D_{m}=C_{\ell m}$ is given by
 \be
 C_{\ell \ell}=-h_\ell  \,   ,  \qquad   C_{\ell,\ell+1}=C_{\ell+1,\ell}=1
 \ee
   Given a cone $\sigma_\ell$, we define the dual cone $\sigma^*_\ell$ as a set of vectors $ v^* \in  \mathbb{R}^2$ such that $v^*\cdot w >0$   $ \forall  w \in \sigma_\ell$.
Equivalently, the generators $(v_{\ell+1}^*,-v_\ell^*)$ of the dual cone $\sigma^*_\ell$  are defined by the conditions
 \be
v^*_\ell\cdot v_\ell=0 \, , \qquad    v^{*1}_\ell  v_\ell^2  -    v_\ell^1  v^{*2}_\ell  >0
\ee
For a vector $v_\ell=(v_\ell^1,v_\ell^2)^T$ one finds the dual vector $v^*_\ell=(v_\ell^2,-v_\ell^1)^T$.

  (\ref{constraint0}) leads to the $\chi-2$ equivalences
\be
\forall \lambda \in \mathbb{C}^* \qquad (y_1,\ldots , y_\chi) \sim (\lambda^{C_{s 1}} y_1, \ldots , \lambda^{C_{s \chi}} y_\chi) , \qquad s=1,\ldots \chi-2
\label{equivalence0}
\ee
  To each dual cone $\sigma^*_\ell$ one can associate a chart $U_\ell$ isomorphic to $\mathbb{C}^2$. Local coordinates in these charts can be taken to be
    \be
    z_1^\ell  =\prod_{\ell'=1}^\chi y_{\ell'}^{v_{\ell+1}^* \cdot v_{\ell'} }    \,  , \qquad     z_2^\ell  =\prod_{\ell'=1}^\chi y_{\ell'}^{-v_{\ell}^* \cdot v_{\ell'} }
    \ee
    Using (\ref{constraint0}) it is easy to see that $z_a^\ell$ are invariant under the action (\ref{equivalence0}).
   \par We introduce a $(\mathbb{C}^*)^2$ action acting on the homogenous coordinates $y_\ell$ as
   \be
   y_1 \to  e^{ \epsilon_1}\,  y_1  \, , \qquad y_2 \to  e^{ \epsilon_2}\,  y_2 \, , \qquad y_{\ell>2} \to    y_\ell  \, .
   \ee
  The action on the local coordinates can then be written as
      \be
  \qquad  z_a^\ell \to  e^{ \epsilon_a^\ell}\,  z_a^\ell
  \ee
  with
  \be
 ( \epsilon_1^\ell,\epsilon_2^\ell)= ( v^{* \, a}_{\ell+1} \, \epsilon_a  ~,\,  -v^{*\, a}_{\ell} \, \epsilon_a  ) = (v_{\ell+1}^2\epsilon_1-v_{\ell+1}^1\epsilon_2,-v_{\ell}^2\epsilon_1+v_{\ell}^1\epsilon_2)  \label{epsilonell}
  \ee
\par The origin  of a patch $U_\ell$  $(z_1^\ell, z_2^\ell)=0$ is invariant under the toric action. We denote this fixed point as $P_\ell$ and in terms of the global coordinates it can be written as $(y_\ell,y_{\ell+1})=0$. Note that every divisor $D_\ell$ contains two fixed points, namely $P_{\ell-1}$ and $P_\ell$.
\par Taking
  \bea
  v_1 &=& (^0_1) \qquad \qquad   ~~~~~~v_2=(^1_0)\nn\\
    v^*_1 &=& (1,0) \qquad \qquad   -v^*_2=(0,1)
  \eea
 one  finds
 \be
 ( \epsilon_1^1,\epsilon_2^1)=( \epsilon_1,\epsilon_2)
 \ee
 The remaining $\epsilon_a^\ell$  can be found from the recursive relations
 \be
 (\epsilon_1^{\ell+1},\epsilon_2^{\ell+1} ) =( h_{\ell+1}\, \epsilon_1^\ell+\epsilon_2^\ell,-\epsilon_1^\ell)  \label{rece}
 \ee
 following from (\ref{constraint0}).

 \begin{remark} \label{numberofregs} An important remark is that for any compact toric variety there are two and only two patches with $\epsilon_1^{\ell}\epsilon_2^{\ell}>0$.
 \end{remark}
 Indeed, as it follows from (\ref{epsilonell}) the signs of $\epsilon_1^{\ell}, \epsilon_2^{\ell}$ depend only on which side of the line $v^2 \epsilon_1 - v^1 \epsilon_2$ the vectors $v_{\ell+1},v_{\ell}$ lie. Since the cones are convex they cover the whole $\mathbb{R}^2$ and $\epsilon_1^{\ell}, \epsilon_2^{\ell}$ cannot be zero, the above statement follows.
  \begin{remark} \label{e1le2l+1}
  $\sign (\epsilon_1^{\ell}\epsilon_2^{\ell+1})=\sign (-(\epsilon_1^{\ell})^2)=-1$.
 \end{remark}

An equivariant two form $w$ on $M$ is defined as a form satisfying
\be
Q_\xi \, w=dw+i_\xi w=0
\ee
with $i_\xi dz_a=\epsilon_a z_a$ the contraction with respect to the action of the $V$ vector field.
To each divisor $D_\ell$ one can associate a Poincar\'e dual equivariant form $w^\ell$ such that
 \be
{1\over 2\pi} \int_{D_\ell}  w^{\ell'}= C_{\ell \ell'}
\ee
The zero form part of $w^\ell$ evaluated at the origin of a patch $U_k$, which we denote as $[w^\ell]_k$, is the equivariant pullback $\iota^*_{P^k \hookrightarrow M}\w^\ell$ of the form $\w^\ell$ via the embedding $P^k \hookrightarrow M$.
The precise form of $[w^\ell]_k$ can be computed using localization.
Let $\alpha$ be an equivariant form. Then, with the help of the localization theorem one can write
\begin{equation} \label{PembX}
    \int_M \alpha \wedge w^\ell = (2 \pi)^2 \sum_{P_k \in M} \frac{\iota^*_{P_k \hookrightarrow M}(\alpha \wedge w^\ell)}{\epsilon_1^{(k)}\epsilon_2^{(k)}}=(2 \pi)^2 \sum_{k=1}^\chi  \frac{[\alpha]_k \, [w^\ell]_k }{\epsilon_1^{(k)}\epsilon_2^{(k)}}
\end{equation}
 The same integral can be computed as an integral  over the dual  divisor $D_\ell$ of the
 equivariant pullback $\iota^*_{D^l \hookrightarrow M} \alpha$ via the embedding $D^l \hookrightarrow M$ . The integral localizes around the fixed points $P_{\ell-1}$ and
 $P_\ell$ intersecting the divisor
\begin{equation}
\begin{aligned}
   \int_M \alpha \wedge w^\ell =   2\pi \int_{D_\ell}   \alpha  =( 2\pi)^2 \left( \frac{ [\alpha]_{\ell-1} }{\epsilon_1^{(\ell-1)}}+
   \frac{ [\alpha]_\ell }{\epsilon_2^{(\ell)}} \right)  \label{PembX2}
   \end{aligned}
\end{equation}
Comparing ( \ref{PembX}) and  ( \ref{PembX2}) one finds
\begin{equation}
   [w^\ell]_k=\iota^*_{P^k \hookrightarrow M}w^\ell= \delta_{k,\ell} \, \epsilon_1^{(\ell)}+   \delta_{k,\ell-1} \, \epsilon_2^{(\ell-1)}
   \ee
    \par Consistently, one can also check that the intersection matrix computed with the localization theorem gives the expected result
\begin{equation}
C_{lm}=\frac{1}{(2\pi)^2}\int w^\ell \wedge w^m =\sum_k \frac{   [w^\ell]_k    [w^m]_k    }{\epsilon_1^{(k)}\epsilon_2^{(k)}}
\end{equation}

\section{The Barnes double gamma function} \label{BDGF}
\setcounter{equation}{0}
The Barnes double gamma function is defined  via analytic continuation to
the whole complex plane of the integral
\be
\log\Gamma_2(x|\epsilon_1,\epsilon_2)  = {d\over ds}\left[  {\Lambda^s\over \Gamma(s) } \int_0^\infty {dt\over t}
{t^s\, e^{-x t} \over
(1-e^{-\epsilon_1 t} ) (1-e^{-\epsilon_2 t} )  }\right]_{s=0}  \label{ge1e2}
\ee
in the region  $  x >0 $ where the integral converges. Using the representation of the logarithm
\be
\log  \left( { \Lambda \over x } \right) ={d\over ds}\left(  {\Lambda^s\over \Gamma(s) } \int_0^\infty {dt\over t}  t^s\,   e^{-x t}  \right)_{s=0} \label{logrep}
\ee
 the double gamma function can be written as an infinite product of zeros or poles according to the domain of definition.
 For example for  $ \epsilon_1, \epsilon_2>0$ writing
  \bea
\log\Gamma_2(x|\epsilon_1,\epsilon_2)  &=& \sum_{i,j=0}^\infty {d\over ds}\left[  {\Lambda^s\over \Gamma(s) } \int_0^\infty {dt\over t}
 t^s\,  e^{-t (x+ i \epsilon_1+j\epsilon_2) }   \right]_{s=0}
= \sum_{i,j=0}^\infty  \log  \left( { \Lambda \over  x+ i \epsilon_1+j \epsilon_2  }\right) \nn
\eea
  one can represent the $\Gamma_2(x)$ function as the infinite product  of poles
\be
\Gamma_2(x|\epsilon_1,\epsilon_2)  =  \prod_{i,j=0}^\infty    \left( { \Lambda \over  x+ i \epsilon_1+j\epsilon_2  }\right) \qquad\qquad  x,\epsilon_1,   \epsilon_2 >0
\label{1looppositiva}\ee
Similarly in the region  $ \epsilon_1>0> \epsilon_2$ one writes
\be
{e^{-x t} \over
(1-e^{-\epsilon_1 t} ) (1-e^{-\epsilon_2 t} )  } ={e^{-x t+\epsilon_2 t } \over
(1-e^{-\epsilon_1 t} ) (1-e^{\epsilon_2 t} )  } =-\sum_{i,j=1} e^{-(x+(i-1) \epsilon_1-j \epsilon_2)t}
\ee
and  the $\Gamma_2(x)$ admits a representation as the infinite product of zeros
\be
\Gamma_2(x|\epsilon_1,\epsilon_2)  =  \prod_{i,j=1}^\infty    \left( {    x+( i-1) \epsilon_1-j \epsilon_2  \over \Lambda }\right) \qquad\qquad  x,\epsilon_1>0> \epsilon_2
\label{1loopnegativa}\ee

\section{Proof of \eqref{idzero2}} \label{proof}

\setcounter{equation}{0}
Let us consider first a variety with $\chi=3$ (for instance, $\mathbb{P}^2$). Every point $\{ k^{\ell} \}$ contributes at most with a simple pole, so in order to compute the residue at the point $a=0$, we have to take only the leading term in the Laurent expansion
of the partition function in each chart. The abstruse duality relates the leading term in each chart to the one obtained from it by flipping the sign of a $k^\ell$. Since every $k^\ell$ appears twice in the product $\prod_{\ell=1}^{3}Z^{\ell}$, once in the $\ell$-th patch and once in the $(\ell-1)$-th patch, according to (\ref{signflip}) one finds
\begin{equation}
{\cal P}_\ell {\rm Res}_{\mathfrak{a}=0} Z^{\chi=3}_{\rm full} ({\bf k} ) =  \sign [\epsilon^{(\ell-1)}_1\epsilon^{\ell}_2]
{\rm Res}_{a=0} Z^{\chi=3}_{\rm full} ({\bf k} )  = - {\rm Res}_{\mathfrak{a}=0} Z^{\chi=3}_{\rm full} ({\bf k} )
\end{equation}
where the last identity  follows from (\ref{rece}).  So we conclude that
 \be
   {\rm Res}_{\mathfrak{a}=0}  \left[   (1+{\cal P}_\ell)  Z^{\chi=3}_{\rm full}( \mathfrak{a}, {\bf k},\epsilon_a,q)  \right]= 0
\ee
and so the residue is also zero. Now let us consider a variety with $\chi=4$ ($\mathbb{F}_n$, for example).
We would like to prove that
\begin{equation} \label{zerores4}
  {\rm Res}_{\mathfrak{a}=0} \left[  (1+{\cal P}_\ell)  (1+{\cal P}_{\ell+1} ) Z^{\chi=4}_{\rm full}(\mathfrak{a}, {\bf k},\epsilon_a,q) \right]= 0
\end{equation}
  If the tuplet$\{ k^{\ell} \}$  contributes as a simple pole, the previous argument is applicable and (\ref{zerores4})  follows. If
  the tuplet contributes a double pole, the residue results from taking the leading terms in the Laurent expansion of three
of the charts and one subleading term. If the subleading term is taken from a chart different from $\ell$,
it is unaffected by the sign flips ${\cal P}_\ell $ or ${\cal P}_{\ell+1}$ and the identity (\ref{zerores4}) follows.

 Finally, let us consider the term  where the subleading contribution comes from the $\ell$-th patch. The flipping of $k^{\ell}$ and $k^{\ell+1}$ affects the subleading contribution. Let us note that if $\epsilon_1^{\ell} \epsilon_2^{\ell}>0$, the $\ell$-th patch contributes as a regular point $(\mathfrak{a}^{\ell})^0$ and hence the subleading term is an odd function of $\mathfrak{a}^{\ell}$. If $\epsilon_1^{\ell} \epsilon_2^{\ell}<0$ the patch contributes as a simple pole and so the subleading term is an even function of $\mathfrak{a}^{\ell}$. All together, one can say that
\begin{equation}
  {\cal P}_{\ell} {\cal P}_{\ell+1}  \mathrm{Sublead} \, Z_{\mathbb{C}^2}^{\ell}  = - \sign( \epsilon_1^{\ell} \epsilon_2^{\ell}) \mathrm{Sublead} \, Z_{\rm \mathbb{C}^2}^{\ell}
\end{equation}
Therefore one finds
\begin{equation}
   (1+{\cal P}_\ell {\cal P}_{\ell+1} ) Z^{\chi=4}_{\rm full}  =\left[1-\sign( \epsilon_1^{\ell-1} \epsilon_1^{\ell} \epsilon_2^{\ell} \epsilon_2^{\ell+1}) \right] Z^{\chi=4}_{\rm full}=0 ,
\end{equation}
\begin{equation}
 ({\cal P}_\ell+ {\cal P}_{\ell+1} ) Z^{\chi=4}_{\rm full} ={\cal P}_\ell  \left[1+{\cal P}_\ell {\cal P}_{\ell+1} \right]  Z^{\chi=4}_{\rm full}=0  .
\end{equation}
  Similar manipulations hold for $\chi>4$, with $\chi-2$ sign flips.

\section{Klyachko's classification of sheaves}
\setcounter{equation}{0} \label{Klyachko}
Here we follow the identification between the fluxes and the positions of the jumps in Klyachko's filtrations first suggested in \cite{Bershtein:2015xfa} (see their Appendix A).
\par According to \cite{Klyachko1,Klyachko2} an equivariant reflexive sheaf on a smooth toric variety $M$ can be defined by a tuple of $\chi$ non increasing filtrations of vector spaces $E_i^{(\ell)}$, $\ell \in \{1 \ldots \chi\}$
\begin{equation}\label{filtration}
\mathbb{C}^N = E^\ell_{-\infty}  \ldots \supseteq E_i^{(\ell)} \supseteq E_{i+1}^{(\ell)} \supseteq   \ldots E^\ell_{\infty} = 0
\end{equation}
\par A sheaf is stable if and only if for any proper subspace $F \subset E=\mathbb{C}^N$ the following inequality holds \cite{KnutsonSharp}
\begin{equation} \label{genstab}
\frac{1}{\dim F} \sum_{i \in \mathbb{Z}} \sum_{\ell=1}^{\chi} i \dim \left(\frac{F \cap E^{(\ell)}_i}{F \cap E^{(\ell)}_{i+1}} \right)\int_M w_\ell\wedge \omega <    \frac{1}{\dim E} \sum_{i \in \mathbb{Z}} \sum_{\ell=1}^{\chi} i \dim \left(\frac{E^{(\ell)}_i}{E^{(\ell)}_{i+1}} \right)\int_M w_\ell\wedge \omega,
\end{equation}
where $w_\ell$ is a form dual to the divisor $D^\ell$ and $\omega$ is the Kahler form. A semistable sheaf is defined by the non strict inequality (\ref{genstab}). A strictly semistable sheaf is semistable but not stable.
 Indeed, a gauge bundle ${\cal E}$ associated to the spaces $E_i^{(\ell)}$ is stable if for any equivariant sub-bundle ${\cal F}$ associated to the induced spaces $F \cap E_i^{(\ell)}$ (\ref{genstab}) holds, or equivalently if the slope of the equivariant sub-bundle is smaller than that of the slope of the bundle itself 
\be
\mu({\cal F}) < \mu({\cal E}) .
\ee
    To understand the content of the stability conditions (\ref{genstab}) it is convenient to  introduce the tuplets of integers  $\{ k^{(\ell)}_\alpha \}$  describing the  positions along the i-line in the $\ell^{\rm th}$-sequence of the jumps from $\mathbb{C}^{\alpha}$ to $\mathbb{C}^{\alpha-1}$. Here for simplicity we use the shift symmetry to set the location of the first jump in each sequence at the origin, i.e.  $k^{(\ell)}_N=0$.

    \subsection{$SU(2)$ case}

 In this simple case there are at most two jumps in a filtration with one one-dimensional intermediate space, which we will denote as $P^{(\ell)}$.  The only non trivial choice for $F$ is $F = P^{(\ell')}$ for some $\ell'$.  Let us first take the sheaves characterised by a tuplet of filtrations with two jumps and $P^{(\ell)}\neq P^{(\ell')}$ for all $\ell \neq \ell'$. Then (\ref{genstab}) reduces to
\begin{equation} \label{494}
\forall \, \ell'   \qquad     \beta_{\ell'} \, k^{\ell'}  \leq  \ft12 \sum_{\ell=1}^\chi \beta_\ell \, k^\ell
\end{equation}
with
\be
\beta_\ell ={1\over 2\pi} \int_M w\wedge w_\ell
\ee
If the inequalities (\ref{494}) are strictly true for any $\ell'$, then  (\ref{genstab}) holds for any $F$ and the sheaf is stable.

   We notice that if the choice of the subspaces $P^{(\ell)}$ is more degenerated, then (\ref{494}) is not a sufficient condition for a sheaf to be stable. For example, if $P^{(\ell)}= P^{(\ell')}$ for any  $\ell,\ell'\in \{1,\ldots, \chi\}$, then (\ref{genstab}) is always false for $F=P^{(\ell)}$, even if (\ref{494}) is still true. It means that the corresponding sheaf is unstable. Therefore we see that (\ref{494}) is only a necessary (but not sufficient) condition for a sheaf defined by the tuplet of filtrations with the positions of the jumps $\{ k^\ell \}$ to be stable.

 In the same way one can see that (\ref{semist}) guarantees that there are some strictly semistable sheaves among all of the sheaves corresponding to a tuplet of filtrations with the jumps at $\{k^\ell \}$.
If both (\ref{st}) and (\ref{semist}) are not satisfied then the non strict version of (\ref{genstab}) cannot be satisfied and so (\ref{unst}) is a sufficient condition for all the corresponding sheaves to be unstable.
 Taking into account also the filtrations with one jump from $\mathbb{C}^2$ directly to $0$ one will end up with the same inequalities (\ref{st} - \ref{unst}) with the corresponding $k^{(\ell)}=0$.

       \subsection{$SU(3)$ case}
    
     Now let us consider the SU(3) gauge theory. The information about the subspaces of the filtrations relevant for the stability conditions in this case can be represented by points and lines in the projective plane. Indeed, two-dimensional subspaces of $\mathbb{C}^3$ can either coincide or intersect at a one-dimensional space; a one-dimensional space either lies in a two-dimensional space or has no non-trivial intersection with it;  for any two non-coinciding one-dimensional spaces there exist one and only one two-dimensional subspace, which include them both (which is their direct sum). Lines and dots on the projective plane have similar properties. The projective plane is required to avoid the existence of parallel lines. The choices for $F$ in this case are given by one and two-dimensional subspaces of $\mathbb{C}^3$, represented by points and lines respectively.
    
    To each divisor $D_\ell$ in the toric manifold one can associated a line $L^{(\ell)} $. We denote by $P^{(\ell)} \in L^{(\ell)}$ a generic point in this line and by $V_{\ell \ell'}$ a vertex at the intersection of the lines. For $\mathbb{P}^2$ the most non-degenerated choice of the subspaces corresponds to the diagram drawn in the following picture
\par
\begin{center}
\begin{tikzpicture}[scale=3]
        \filldraw[black] (0.2,0) circle (0.5pt);
        \filldraw[black] (0.3,0.6) circle (0.5pt);
        \filldraw[black] (0.9,0.2) circle (0.5pt);
        \node[] at (-0.05,-0.2) {$ L^{(1)}$};
        \node[] at (0.3,1.1) {$L^{(2)}$};
        \node[] at (1.25,-0.05) {$ L^{(3)}$};
        \node[] at (0.2,.65) {$P^{(1)}$};
        \node[] at (1.05,.25) {$P^{(2)}$};
        \node[] at (.2,-.1) {$P^{(3)}$};
         \draw[black, thick] (-0.2,0)--(1.2,0);
         \draw[black, thick] (-0.1,-0.2)--(.6,1.2);
         \draw[black, thick] (0.4,1.2)--(1.1,-0.2);
\end{tikzpicture}
    
\end{center}

There are now four relevant choices for $F$:

\begin{itemize}

\item{ $F=P^{(m)}$: A point intersects itself and the line that it contains it, so $F \cap  L^{(m)} = F \cap P^{(m)}=F$ leading to
\be
\forall m \qquad k_1^{(m)} \leq {1\over 3} \sum_{\ell=1}^3(k_1^{(\ell)}+k_2^{(\ell)})
\ee
 
 }

\item{ $F=L^{(m)} \cap L^{(n)}=V_{mn} $:  A vertex intersects two lines $F = F \cap  L^{(m)}  = F \cap  L^{(n)}  $ leading to 
\be
\forall m\neq n  \qquad k_2^{(m)}+k_2^{(n)} \leq {1\over 3} \sum_{\ell=1}^3(k_1^{(\ell)}+k_2^{(\ell)})
\ee

 }

\item{ $F=L^{(m)}$: A line intersects itself on a line, a point contained in it on a point, and any other line on a vertex, so: 
  $F \cap  L^{(m)} = F$, $ F \cap  P^{(m)}=P^{(m)}$, $F \cap L^{(n)}=V_{mn}$, leading to
   
\be
\forall m  \qquad \ft12(k_1^{(m)}+ k_2^{(1)}+k_2^{(2)}+k_2^{(3)})  \leq {1\over 3} \sum_{\ell=1}^3(k_1^{(\ell)}+k_2^{(\ell)})
\ee

 }

\item{ $F=P^{(m)} \oplus P^{(n)} $: A line connecting two points intersects the two points and all the lines contained on a point, 
so
  $ F \cap  L^{(m)}=F \cap  P^{(m)} =P^{(m)}$, $ F \cap  L^{(n)}=F \cap  P^{(n)} =P^{(n)}$,  $ F \cap L^{(3)} =\tilde{P}^{(p)}$, leading to  
\be
\forall m\neq n\neq p  \qquad \ft12( k_1^{(m)}+k_1^{(n)} +k_2^{(p)})  \leq {1\over 3} \sum_{\ell=1}^3(k_1^{(\ell)}+k_2^{(\ell)})
\ee

}

\end{itemize}
   
  Like in the SU(2) case, the resulting 12 conditions are necessary but not sufficient conditions for stability, because they are given by the most non-degenerated choice of the subspaces of the filtrations.  
  
\end{appendix}

\providecommand{\href}[2]{#2}\begingroup\raggedright\endgroup

\end{document}